\documentclass[11pt]{article}
\usepackage{caption,amsmath,graphicx,url,amssymb,latexsym,epsfig,tabularx,setspace,multirow,threeparttable,longtable,pdflscape,tabu,appendix} 
\usepackage[round, sort, numbers, authoryear]{natbib}
\usepackage[margin=1in]{geometry}

\begin{document}
\captionsetup{width=\textwidth}

\begin{titlepage}
\centering

\textbf{\Large Assessing Researcher Interdisciplinarity: A Case Study of the University of Hawaii NASA Astrobiology Institute}\\
\Large Michael Gowanlock \& Rich Gazan\\
\vspace{2cm}

\raggedright
M. Gowanlock\\
Department of Information \& Computer Sciences and University of Hawaii NASA Astrobiology Institute, University of Hawaii\\ POST 310, 1680 East-West Road, Honolulu, HI 96822, USA\\
gowanloc@hawaii.edu\\
\vspace{1cm}
R. Gazan\\
Department of Information \& Computer Sciences, Library \& Information Science Program and University of Hawaii NASA Astrobiology Institute, University of Hawaii, Hamilton Library 2H, 2550 McCarthy Mall, Honolulu, HI 96822, USA\\
gazan@hawaii.edu\\

\end{titlepage}


\begin{abstract} 
In this study, we combine bibliometric techniques with a machine learning algorithm, the sequential Information Bottleneck, to assess the interdisciplinarity of research produced by the University of Hawaii NASA Astrobiology Institute (UHNAI).  In particular, we cluster abstract data to evaluate Thomson Reuters Web of Knowledge subject categories as descriptive labels for astrobiology documents, assess individual researcher interdisciplinarity, and determine where collaboration opportunities might occur.  We find that the majority of the UHNAI team is engaged in interdisciplinary research, and suggest that our method could be applied to additional NASA Astrobiology Institute teams in particular, or other interdisciplinary research teams more broadly, to identify and facilitate collaboration opportunities. 

\vspace{1cm}

\textbf{Keywords: Astrobiology, Bibliometrics, Information bottleneck method, Interdisciplinary science, Machine learning, Text mining}
\end{abstract}

\section{Introduction} Astrobiology, the study of the origin,
evolution, distribution, and future of life in the universe, is a
relatively new field comprised of researchers from a range of scientific
disciplines.  Apart from its sublime object of study, astrobiology has
been identified as a field that can integrate diverse sciences
\citep{JamesT2003347}, provide a tangible target for interdisciplinary
science education \citep{Charles2002263}, and provide a pathway to adult
science literacy \citep{Oliver2007716}.  Many of the field's core
questions require knowledge from multiple disciplines to be harvested,
integrated and applied outside of their source domains, and as such,
astrobiology is inherently interdisciplinary.  For example, the
University of Hawaii NASA Astrobiology Institute (UHNAI) studies the
origin of water in the solar system and beyond, in the context of
understanding the origins of life. Astronomers, chemists, geologists,
oceanographers and biologists work together to study data from meteorite
fragments to comets to the interstellar medium to address the question
of where else in the universe water, and thus life, might be found.
Without collaboration across disciplinary boundaries to interpret
often-scarce data, important questions in astrobiology will remain
incompletely addressed. Developing a method to identify, measure and
catalyze interdisciplinary work in the astrobiology research environment
is the goal of this paper.

One of the benefits of a broad-based research community is that new
developments in astrobiology occur fairly frequently.  The downside is
that researchers must stay abreast of these numerous developments both
inside and outside of their home fields.  As new astrobiology research
findings are reported, the considerable effort involved in finding,
evaluating and integrating them indicates a need for a better
understanding of how findings in one field might inform others, and to
identify potential collaboration opportunities between individual
researchers working on similar questions.

Our previous example suggested that knowledge from multiple disciplines is required to understand the origin of water to answer questions regarding the origin of life. Satisfactorily understanding the research record of scientists that work in this area requires measuring interdisciplinarity on an acute scale. Following \citet{van_leeuwen_2007}, we distinguish between a top-down bibliometric approach, where large-scale trends at the highest levels of publication aggregation are considered (such as the research output of a country or university), and prefer a bottom-up approach, where we analyze individual documents and the papers they cite.  We harvest each astrobiologist's publication data by comparing NASA Astrobiology Institute annual reports, where publications are systematically documented, with the researchers' Websites and CVs, and further verify the data by browsing the author indexes of each database to identify name variations, to represent the research output of each astrobiologist. 

A common method used to examine the potential of collaboration across
disciplinary boundaries is to interview domain experts, but this method
suffers from several limitations, such as sample size and
subjectivity problems \citep{springerlink:10.1007/s11192-010-0318-1}. Furthermore,
given that the subject matter of astrobiology spans many disciplines,
meaningful analysis of the responses would require the knowledge of an
astrobiology polymath.  After considering these limitations, we suggest
that measuring interdisciplinarity should be guided by one or more
individuals versed in astrobiology, but whose expertise need not span
all of its constituent disciplines.  Therefore, an unsupervised approach
is optimal as such methods can find trends in data without prior knowledge of its structure.

As of 2011, the NASA Astrobiology Institute is comprised of 14 teams
spanning ten universities in addition to NASA Ames, Goddard, and the Jet
Propulsion Laboratory.  While a cross-team analysis is beyond the scope
of this paper, we suggest that our method for measuring researcher
interdisciplinarity at UHNAI could be extended to other NASA Astrobiology Institute teams, and
to scientific collaborations more broadly. Furthermore, our method
suggests where collaborations might productively occur, and allows us to
better understand the nature of interdisciplinary scientific discovery.

In this pilot study, we investigate the use of an unsupervised machine
learning clustering technique, the sequential Information Bottleneck
(sIB) \citep{Slonim:2002:UDC:564376.564401} to aid in measuring
researcher interdisciplinarity.  Furthermore, we assess the extent to which Journal Subject Categories from the Thomson Reuters Web of Knowledge database suite are sufficient for labelling astrobiology documents.   The
clustering and classification of text allow interdisciplinary analysis
that 1) describes collaboration and the integration of knowledge and 2)
draws conclusions that are useful to astrobiology researchers by
uncovering the underlying structure of research tracks.  The results of
this pilot study will serve to guide a subsequent investigation that
will identify collaboration opportunities and measure the disciplinary
roots across the entire NASA Astrobiology Institute.

Researchers in astrobiology tend to be comfortable speaking in the
language of multiple scientific disciplines.  As suggested in \cite
{springerlink:10.1007978-94-007-1658-2}, these interdisciplinary
researchers are somewhat isolated from their counterparts in other
academic departments.  The multidisciplinary context given by
astrobiology affords an excellent opportunity to examine the methods
used to study researcher interdisciplinarity and knowledge integration. Furthermore, we propose an iterative process to identify specific publications that bridge diverse fields, to facilitate interdisciplinary collaborations and ease the cognitive load of a single researcher who wishes to integrate knowledge from multiple disciplines.

\section{Background} 
Research that occurs at the intersection between
disciplines is thought to lead to great advances in science
\citep{s11192-008-2197-2}.  Many funding agencies exist specifically to
support and encourage interdisciplinary research; the U.S. National
Science Foundation's interdisciplinary research efforts span all of
their divisions and directorates \citep{NSF_IDR}.  For example, some authors measuring
interdisciplinarity lament that there is not enough coverage of the
societal causes for climate change
\citep{springerlink:10.1007-s11192-011-0356-3} as described in the 
Intergovernmental Panel on Climate Change (IPCC) literature.  In this
specific case, measuring both the disciplinary diversity and the
integration of knowledge is of paramount importance to ensure that
future IPCC reports include appropriate factors.  A cynical disposition
to this problem is eloquently stated in
\cite{springerlink:10.1023/A:1004706019826}: ``The world has problems,
but universities have departments."

Many important terms in this work have so far been discussed without
qualification.  The term interdisciplinary tends to be tacitly understood
by researchers, with no consensus definition.  We adopt the definition
suggested by \citet{springerlink:10.1007s11192-007-1700-5}, which
followed the definition given by the
\citet{NatAcademies}: interdisciplinary research requires an integration
of concepts, theories, techniques and/or data from two or more bodies of
specialized knowledge.  Multidisciplinary research may incorporate
elements of other bodies of specialized knowledge, but without
interdisciplinary synthesis \citep{Wagner201114} that leads to research
that is greater than the sum of its parts.

Despite the increase in claimed interdisciplinarity, traditional
indicators are of questionable value in assessing and quantifying
interdisciplinary research \citep{springerlink:10.1023A:1010529114941}.
Additionally, policies regarding interdisciplinarity are often based
more on conventional wisdom than empirical studies
\citep{springerlink:10.1007-s11192-009-0041-y}. The usefulness of
bibliometric indicators depends critically on the level at which we wish
to understand the integrative process.  For example, funding agencies
may only require high-level publication co-authorship and collaboration
statistics, describing the research performed by their grantees and the
diversity of their home disciplines, but not addressing the essential
aspect of synthesis.  When there is no mechanism to identify, measure
and encourage these points of intellectual crossover, there is no way to
quantify the extent to which interdisciplinary science is taking place.

Top-down approaches have been used to map scientific literature (for
example, see \citet{springerlink:10.1007/s11192-005-0255-6}), and often
represent broad areas of science with Web of Knowledge (WoK) subject
categories (SCs). For example, \citet{vanRaan2002611} and
\citet{springerlink:10.1007s11192-007-1700-5} used SCs in their
methodology to measure interdisciplinarity.  In these studies, SCs have
been employed as de facto disciplinary boundaries, and as a benchmark to
measure how much a given author, journal or research area crosses
scientific fields.  Unfortunately, low-level conclusions that might
inform potentially productive individual collaborations cannot be made
when relying on these top-down approaches, as they focus on past outputs
rather than future integration.  Conversely, bottom-up bibliometric approaches incorporate the authors'
own words, in free-text fields such as: titles, abstracts,
keywords\footnote{Keywords are not always a free-text field.} and the
full text of a document.  Clustering bibliometric data at this level can describe the structure of a researcher, journal or an entire field, and suggest productive future directions.  Comparing the bottom-up clustered output with the top-down approach of SCs for astrobiology publications yields an indication of the effectiveness of SCs as document labels for works in this interdisciplinary domain. \citet{springerlink:10.1007/BF02458392} describes how a citation analysis
can serve as a ``radioactive trace'' of research impacts.  One limitation
of cluster analysis is that ``...precise disciplinary divisions are not
obtained, rendering inter-cluster links misleading''
\citep{springerlink:10.1007-s11192-009-0121-z}, but
\citet{springerlink:10.1007/s11192-009-0051-9} propose a methodology to
identify emerging ``research fronts'', highly cited micro-specialty areas
that transcend existing fields.  Their method requires that the
researcher not presuppose the existence of any research field, but to
rely instead on a comprehensive monitoring of citations to identify
points across disciplines where research interests intersect, echoing
one goal of the present study.  Both top-down and bottom-up approaches
are useful in different applications. A study by
\citet{springerlink:10.1007-s11192-009-0041-y} combines bottom-up and
top-down approches to measure both disciplinary diversity and knowledge
integration.

Measuring scientific output in bibliometric terms requires some degree
of integration and normalization of the publication records of
researchers, which are published in diverse formats, venues and
scholarly traditions. The publication record generally includes data
such as departmental affiliations, keywords, year of publication,
journal, cited references, and the abstract and/or full text of the
publication.  This data can be compared using various bibliometric techniques
to assess interdisciplinary research.  While bibliometric studies tend
to rely on a citation analysis, such an analysis may not be appropriate
for every discipline or field.  For example, a given field may tend to
reference conference proceedings, websites, newspapers, or colloquia
which are not as conducive to a co-citation analysis as journal
articles.  Due to this observation,
\cite{springerlink:10.1007/s11192-010-0275-8} suggests that studying
interdisciplinarity should include publications beyond journal articles.
While we agree with this position, it happens that journals are the
preferred method of communication within the great majority of the
fields that compose the UHNAI team; therefore, the present study is not
hindered by this limitation.

\section{Methodology}\label{methodology} In this section, we outline our method for
measuring interdisciplinary research.  In the previous section we noted
that particular bibliometric indicators are conducive to understanding
research at varying levels.  One of the goals of this research is to
uncover the underlying structure within an astrobiology research team
that undertakes interdisciplinary projects at the macro scale, but may
differ in the extent of interdisciplinary work at the micro level.  To understand the research structure, we examine the abstract text of research publications and employ a method from the field of information theory, the sIB method, to cluster our high dimensional abstract data.

An advantage of using WoK for bibliometric studies is that it provides a
mapping of SCs to each journal.  Given the incommensurability of other
bibliometric data (for example, journals do not agree upon a common set
of keywords), SCs provide a way to compare publications on the journal
level.  In \citet{Zhang2010185}, the authors used a cross-citation analysis
to create seven high-level clusters of related SCs, though their
analysis was somewhat confounded by the ``idiosyncrasy'' that a journal
may be assigned to multiple SCs in WoK.  In
\citet{springerlink:10.1007s11192-007-1700-5}, the authors examine the
references in sets of journal articles gathered from WoK, and relate the
journals to their corresponding SCs.  In this approach, a more diverse
set of SCs that represent a paper derived from its references indicates
a higher degree of interdisciplinarity than a set of similar SCs that
represent a paper.

Using the references of a paper is a reasonable approach to measure
researcher interdisciplinarity.  Analoguous to
\citet{springerlink:10.1007s11192-007-1700-5}, we use the references in
each UHNAI publication.  In particular, we combine all of the abstracts
of all of the references cited by a UHNAI publication, and use these
aggregated abstracts to represent each publication.  In another text mining study \citep{ASI:ASI1181}, employed free-text fields (such as title, keywords
and abstracts) of cited/citing publications in combination with phrase
frequency analysis and phrase clustering analysis to obtain a low-level
understanding of research impact and interdisciplinary research.

In the present study, we focus on the abstracts of cited papers only,
and we do not consider the papers that cite the UHNAI papers.  A major limitation of studying the citations to the UHNAI
papers is that it would require the database to contain those papers that cite a particular work, which varies between disciplines, fields and databases.  The
same is true of those references that are cited in our UHNAI papers.  To obviate this problem, we elect to use the NASA Astrophysics Data System (ADS) to collect the majority of our abstracts, as more UHNAI team publications are covered in this database than any other.  The extensive coverage in ADS ensures that a considerable majority of
papers referenced by the UHNAI team are also within the database. However, previous research has
illustrated how the differences in scientific publication patterns
between fields often require that records from multiple databases be
harvested to encompass the output of interdisciplinary scientific
researchers
\citep[see, for example,][]{springerlink:10.1007/s11192-008-0217-x}.  For UHNAI authors whose
publications were not sufficiently represented in ADS, we used WoK to
obtain their publication data and cited references.  As it turns out,
those authors, and the papers they cite, were highly represented in WoK.
We were able to gauge author coverage in ADS and WoK by consulting the
CV of each UHNAI team member.

In the following subsections, we describe our methods used to achieve the following goals:
\begin{itemize}
\item  Examine whether WoK SCs are sufficient for labelling astrobiology
documents. While we believe SCs are useful in mapping broad scientific research trends in established disciplines, whether they are appropriate for classifying individual publications or interdisciplinary works remains an open question\footnote{The
classification of documents is a requirement for an astrobiology
publication information retrieval system. Our research group is inclined
to create such a system. See \url{http://airframe.ics.hawaii.edu/} for more
information.}.  We cluster a corpus of astrobiology abstracts labelled
with their corresponding conflated SCs (Section~\ref{method_sc}), and assume that if a given cluster is comprised mostly of a single SC, then SCs are a sufficiently accurate classifier.
\item Identify actual and potential instances of interdisciplinary research in astrobiology using conflated SCs (Section~\ref{method_sc}).
\item Identify actual and potential instances of interdisciplinary research and identify potential collaboration opportunities between researchers using aggregated abstracts to represent the research tracks of the UHNAI team (Section~\ref{method_aa}).

\end{itemize} 

\subsection{Text Mining and the Sequential Information Bottleneck (sIB) Method}\label{sib_motivation}
The sIB clustering algorithm \citep{Slonim:2002:UDC:564376.564401} is employed to cluster our datasets
described below.  We chose this clustering method over others because it
has been shown to perform better than other unsupervised clustering methods,
such as k-means \citep{Slonim:2002:UDC:564376.564401}.  Furthermore, the approach should allow us to identify instances of interdisciplinary research by examining the cluster membership of our abstract data without prior knowledge of the data's properties.  It is necessary to use an unsupervised clustering
method because a canonical set of astrobiology documents with which to
train a clustering technique does not exist.     

\subsection{Data Collection} \label{data_collection}
We gather publications by the UHNAI team
members from 2001 until June 2011.  Publications earlier than 2001 were
not collected, as many researchers may not have been engaged in
astrobiology research, and the UHNAI team was not yet
founded\footnote{This is also the year that the journal Astrobiology
began publication.  While astrobiology research was, and continues to be
published in other journals, this indicates that astrobiology research
may not have coalesced as a field prior to 2001.}.  However, we place no
age restrictions on the papers that they cite.

\subsubsection{NASA Astrophysics Data System (ADS)} The ADS has
extensive coverage of astronomy, astrophysics and physics journal
articles, pre-prints and conference proceedings.  We gather the data in
a semi-automated fashion.  Instead of accessing the articles through a
web browser, ADS has a perl script library\footnote{The scripts can be
found here: \url{http://vo.ads.harvard.edu/adswww-lib/}} that can be used to
access parts of the database.  To gather the abstracts and journals of
UHNAI papers, and the abstracts and journals of the papers they cite, we
employed the following procedure:

\begin{itemize} 
\item Ran one of the ADS perl scripts to return a list
of all of the publications for each UHNAI team member.  This returned a
list of ADS bibcodes, which uniquely identify each record in the ADS.
\item Compared these papers with each author's CV to ensure that we did
not collect undesired articles.  For example, we filtered out papers by
authors with the same last name as members on the UHNAI team. 
\item Used the ADS bibcodes to create a script that goes to the URL of the webpage
that lists the references in each UHNAI paper.  We download the
individual webpages. 
\item Created and ran a script to capture all of the ADS bibcodes in each downloaded html webpage. 
\item Used this list of bibcodes to get the abstracts and journals of all of the UHNAI papers
and references therein using the ADS perl scripts. 
\end{itemize}

\subsubsection{Web of Knowledge (WoK)} 
To include the published output of UHNAI authors whose work is
underrepresented in ADS, and to provide a comprehensive portrait of the entire UHNAI
team, we also used WoK to gather abstracts and bibliographic data.  To
our knowledge there is not an API or alternative way to access WoK other
than using a web browser.  To gather this data, we employed the
following procedure: \begin{itemize} 
\item  Created a list of all of the papers authored by UHNAI authors that were not in or underrepresented in
the ADS database. 
\item Manually downloaded the html pages of each record describing each UHNAI publication and references therein. 
\item Created a script that parses the html pages to harvest the abstracts and journal titles.
\end{itemize}

When working with WoK data, it is important to be mindful of the
differences in institutional subscriptions, which include access to
different subsets and date ranges of WoK's constituent databases, and
may affect the results of bibliometric analysis
\citep{jacsocurrentscience,springerlink:10.1007/s11192-009-0118-7}.  Therefore, we provide a
list of the University of Hawaii WoK subscriptions at the time of data collection:

\begin{itemize} \item Web of Science, 1980 - \item Biological Abstracts,
1969 - \item Medline, 1950 - \item Journal Citation Reports Science \&
Social Science editions, 2004 - \end{itemize}

\subsection{WoK Subject Categories and Document Classification}\label{method_sc} 
Having collected the abstracts and
journal names of UHNAI publications and references, we create a dataset
that contains the abstract text and the SC of the associated journal for
each UHNAI publication and the publications they cite.

Many of the SCs of the papers in our dataset were significantly
underrepresented.  Furthermore, as other researchers have encountered
\citep[see, for example,][]{Zhang2010185} some journals in WoK are
assigned multiple SCs, necessitating some conflation into superclusters,
or ``macro-disciplines'' \citep{s11192-008-2197-2}.  We modify the SCs using the following method: 
\begin{itemize} 
\item Journals with a single WoK SC that appears 10 or more times in our dataset uses the assigned WoK SC name. 
\item Journals with a single WoK SC that appears less than 10 times is changed to a broader WoK category (e.g. ``Biochemical Research Methods'' becomes ``Biochemistry \& Molecular Biology''). 
\item Journals with two or more SCs of roughly equivalent weight are assigned a new conflated SC (e.g. ``Astrophysics \& Geophysics''). 
\item Journals with two or more SCs that have a clear primary SC have ``-Multidisciplinary'' appended to the primary name.
\end{itemize}

The ADS system also contains non-journal publications.  In these instances, we manually assigned an appropriate SC to the publication. Table~\ref{tab:sc_mapping} shows the mapping of WoK SCs to our conflated SCs. 

We eliminated those abstracts whose SCs were unique or constituted a very small fraction of the entire dataset.  Furthermore, publications are commonly cited across the UHNAI team; therefore, multiple duplicate abstracts could appear in our dataset.  Duplicate abstracts were removed from the sample.  Once removing the duplicate abstracts, we traced back which databases these abstracts came from to reflect the relative proportion of abstracts obtained through WoK and ADS.  The dataset has 10216 abstracts integrated over 13 conflated SCs.  Table~\ref{tab:SCs} shows the number and fraction of abstracts labeled with each conflated SC.  From Table~\ref{tab:SCs}, we observe that there is a large class imbalance problem, as the Astronomy \& Astrophysics SC contributes 67.68\% of the entire dataset.

\begin{figure}[tbp] \begin{center}
\includegraphics[width=1\textwidth]{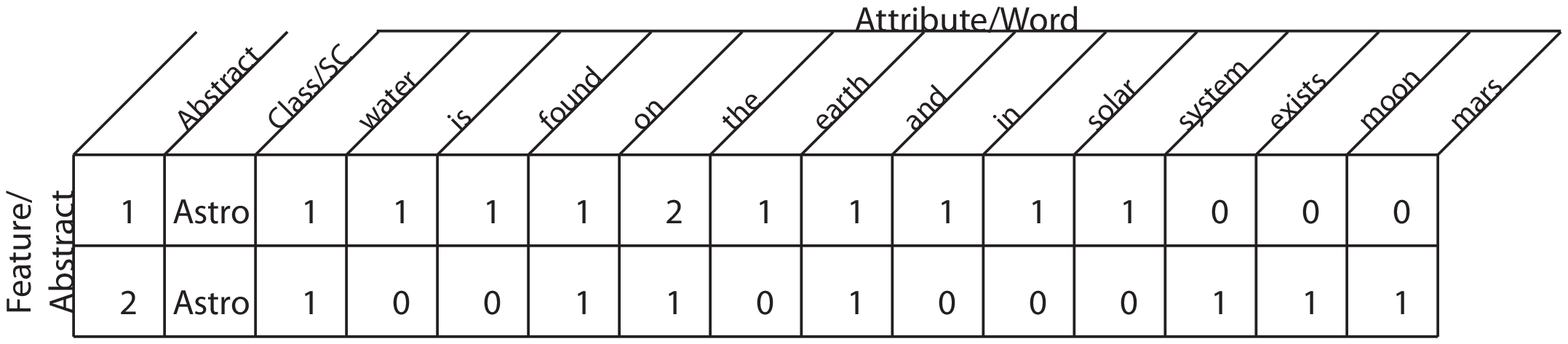} \end{center}
\caption{Depiction of feature vectors as constructed from abstract data.}
\label{feature_vector} \end{figure}

We oversample the minority classes (where a class is a SC), which is every SC but Astronomy \& Astrophysics, to examine if the class imbalance problem significantly affects the resultant clusters. There are a number of methods utilized to oversample minority classes in the field of data mining.  Duplicating the abstracts in the minority classes could potentially result in model
overfitting.  To obviate this problem, we create synthetic data that is similar to the other abstracts within a given SC.  We use the Synthetic Minority Over-sampling Technique (SMOTE) \citep{smote} to produce synthetic feature vectors, where a feature vector (or feature) is a normalized numerical representation of the words that describe each abstract/instance.  For example, consider the following two truncated abstracts:
\begin{enumerate}
\item Water is found on the earth and in the solar system.
\item Water exists on the moon, and Mars.
\end{enumerate}
The two feature vectors of word counts after punctuation is removed is shown in Figure~\ref{feature_vector}.

\begin{figure}[tbp] \begin{center}
\includegraphics[width=0.6\textwidth]{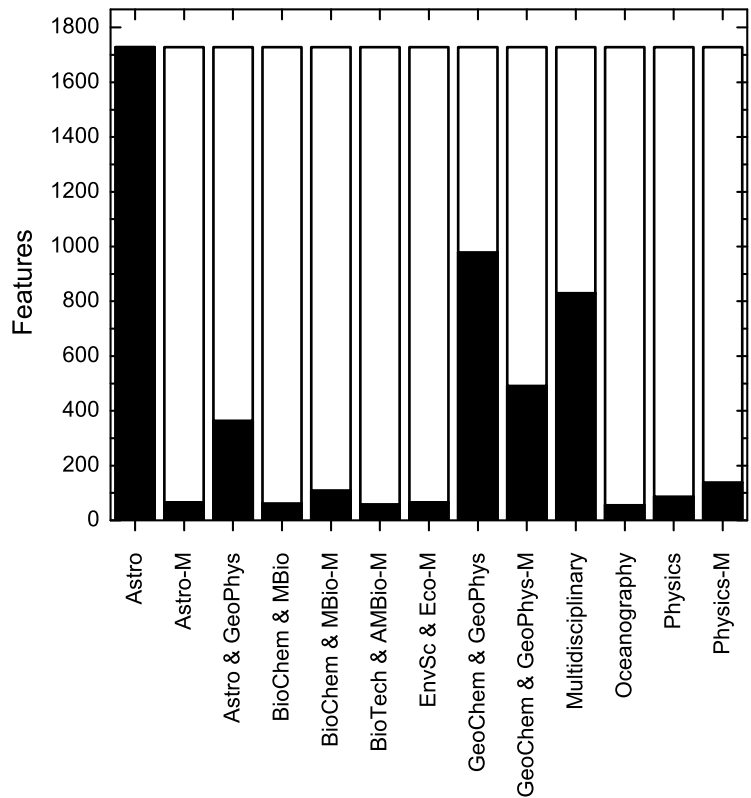} \end{center}
\caption{The distribution of real (black) and synthetic (white) data in the \emph{conflated\_SC\_sampled} dataset.}
\label{real_synthetic_data} \end{figure}

For clustering our abstracts and their corresponding conflated SCs, we create two datasets.  In the first dataset (hereafter \emph{conflated\_SC\_default}), we cluster the dataset as described by Table~\ref{tab:SCs}, without considering the class imbalance problem.   In the second dataset (hereafter \emph{conflated\_SC\_sampled}), we randomly sample without replacement 25\% of the features contained within the Astronomy \& Astrophysics SC and every feature in the minority SCs three times. We use SMOTE to create synthetic feature vectors for the minority SCs such that each SC is represented by the same number of features.  A visual representation of the distribution of real and synthetic data is shown in Figure~\ref{real_synthetic_data}.

\subsection{Text Mining Aggregated Abstracts}\label{method_aa} We create
a dataset of aggregated abstracts (hereafter \emph{aggregated\_abstracts}) for the purposes of representing each
UHNAI publication.  The dataset contains 731 publications by the UHNAI
team.  Table~\ref{tab:authors} shows the team members and their associated home disciplines.  Each publication is represented by its own abstract and the
abstract of each cited publication.  We aggregate all of these abstracts in a single feature vector
to represent each UHNAI publication.  Non-journal publications such as book chapters, conference proceedings and dissertations were included in the dataset, although they constitute
a very small fraction of the total publications.  A majority of the abstracts in the \emph{aggregated\_abstracts} dataset are the same as the ones in the \emph{conflated\_SC\_default} and \emph{conflated\_SC\_sampled} datasets.

We estimate the completeness of the aggregated abstracts, which is defined as the fraction of abstracts harvested out of the total number of citations in a UHNAI publication. For example, if an individual UHNAI publication contains 20 referenced citations, and 15 corresponding cited abstracts were harvested, then the aggregated abstract is 75\% complete.  We randomly sampled ($N$=100) abstracts from the 731 in the \emph{aggregated\_abstracts} dataset.  We find that the average completeness for the aggregated abstracts in this sample is 74.3\%, as shown in Table~\ref{tab:AAs}.  Therefore, we expect that on average our aggregated abstracts in the \emph{aggregated\_abstracts} dataset are $\sim$74\% complete.  Interestingly, the UHNAI publications harvested from ADS have a higher degree of completeness than those abstracts harvested from WoK.

\subsection{Preprocessing of the Datasets}\label{preprocessing} 

We preprocessed the \emph{conflated\_SC\_default}, \emph{conflated\_SC\_sampled} and \emph{aggregated\_abstracts} datasets in the same manner.  Our preprocessing of the datasets included converting uppercase words to lowercase, and ignoring non-alphabetical characters.  We stemmed the words using the Porter stemming algorithm \citep{porter} to ensure that related words were not duplicated in the datasets.  We created a
stopword list to remove formatting tags, and other non-content-bearing
terms.  We selected words which had a minimum frequency of
12, integrated over the entire datasets, resulting in a total of $\sim$4000 words in each dataset.  Most of our preprocessing was performed in WEKA \citep{weka}, and the sIB method was also executed in
this environment.

\begin{figure}[tbp] \begin{center}
\includegraphics[width=0.7\textwidth]{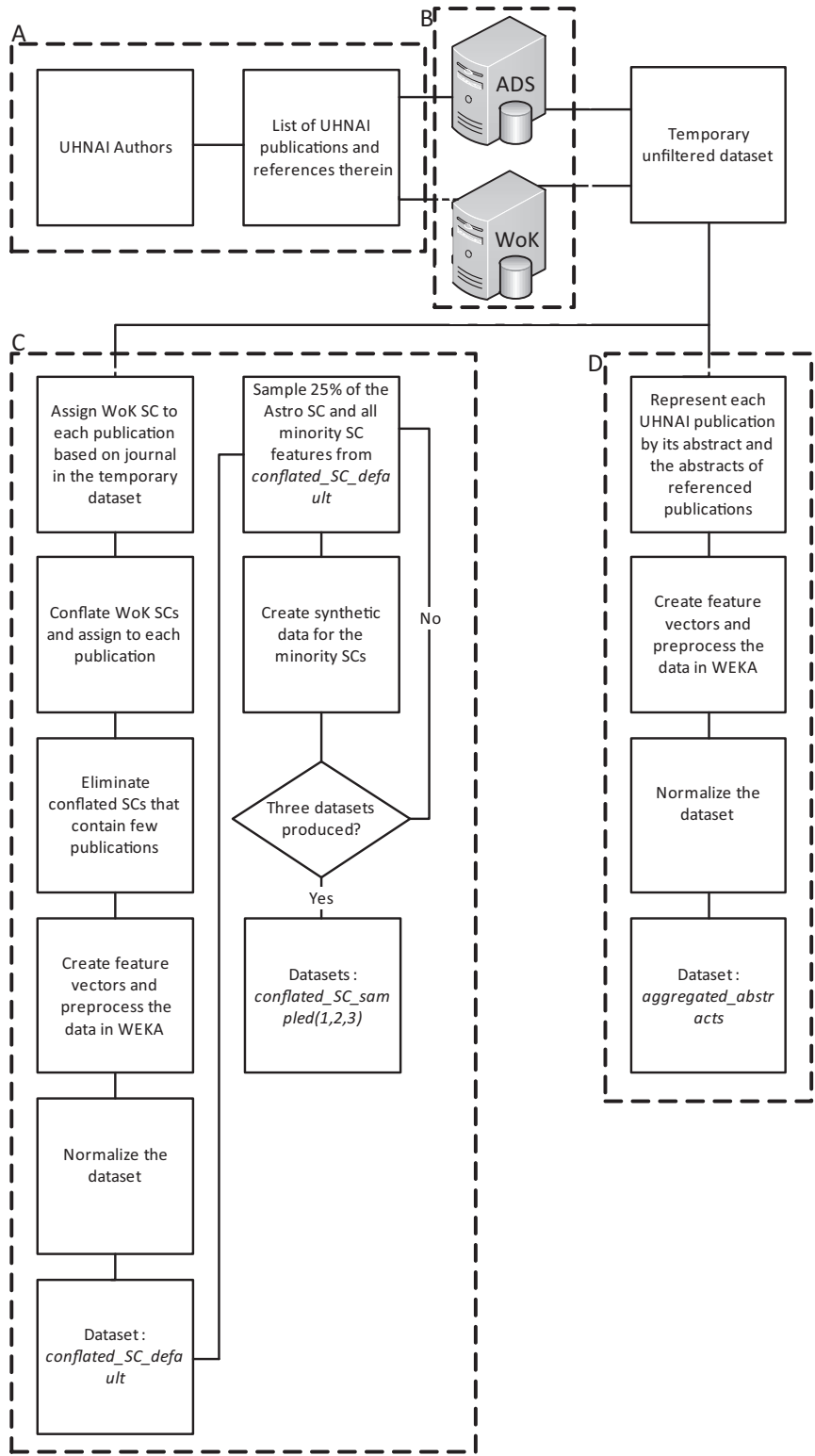} \end{center}
\caption{An overview of the steps in our methodology.  The UHNAI authors and publications to be harvested are are shown as region A and outlined in Section~\ref{methodology}.  The motivation for using the sIB is discussed in Section~\ref{sib_motivation}.  The data collection procedure is discussed in Section~\ref{data_collection} and is described by region B.   The method for creating our conflated subject categories is shown as region C and discussed in Sections~\ref{method_sc} and~\ref{preprocessing}.  The method for creating our aggregated abstracts is shown as region D and discussed in Sections~\ref{method_aa} and~\ref{preprocessing}.  }
\label{method_diagram} \end{figure}

We normalize each feature vector in our datasets. Each feature is described by the term
frequency of each word found in the $\sim$4000 words distilled from
their respective datasets.  We normalize the sum of each feature vector
to 1.  In the case of the aggregated abstracts, some feature vectors
will be much shorter or longer than others, as there is a large range of
abstract sizes, and number of references within a given publication.  If
we did not normalize the term frequencies, then instances with high word
or low word counts may cluster together.  Such clusters would be less
revealing of the content of the documents themselves.  Figure~\ref{method_diagram} reiterates the steps employed to construct our datasets.

\subsection{Limitations} There are several limitations to our study.
First, some of the papers were authored by multiple members of the UHNAI
team.  In this case, we assigned the abstract data to the first-listed
author on the paper, thereby not fully characterizing the research
contribution of the non-primary authors.  Otherwise, having multiple
labels on the same document would inadvertently oversample those documents with
multiple UHNAI authors.  Also, there is a minor discrepancy between the
abstracts gathered in ADS and WoK; ADS contains abstracts from
non-journal sources, whereas WoK does to a lesser extent for the researchers studied here.  The vast majority of our data was from journal articles; therefore, we do not expect this to have a
significant, if any negative impact on our study.  WoK maps multiple SCs
to a single journal.  While we need to conflate the SCs in order to
compare them to clusters (in the \emph{conflated\_SC\_default} and \emph{conflated\_SC\_sampled} datasets), the aggregation procedure undermines the
fundamental function of SCs. Furthermore, we reduced the total number of conflated SCs to 13 which may have a negative effect on our ability to assess interdisciplinary research.

\section{Results} In this section we present the results of our text
mining experiments.  For the purposes of this paper, where our goal is to identify actual and potential instances of interdisciplinary research in astrobiology, a meaningful cluster relationship is one where papers from two or more SCs cluster together, or when researchers from different fields have the aggregated abstracts of their papers cluster together.  Our present and future work is focused on these heterogeneous clusters, however our method could be used for a variety of purposes, each with a different corresponding indicator of interest.  For example, a research team wishing to demonstrate its uniqueness within a collaboration might highlight its work being represented as a relatively homogenous cluster, with its dominant SC not found in other clusters.  A group seeking to align or connect itself with researchers in a particular area might target clusters where their work and that of their target domain co-exist.

\subsection{Subject Categories as Document Labels}\label{results_SC_DL}

\begin{figure}[tbp] \begin{center}
\includegraphics[width=0.5\textwidth]{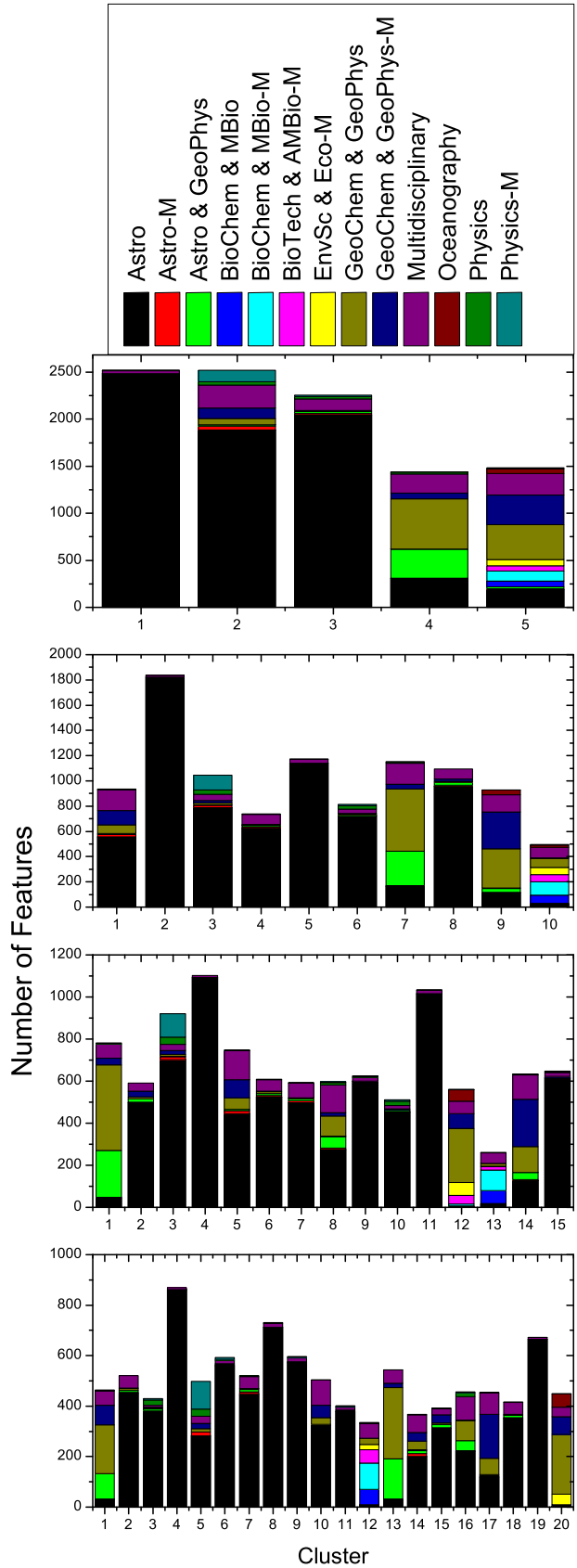} \end{center}
\caption{The results of clustering the \emph{conflated\_SC\_default} dataset. Results are given for 5, 10, 15 and 20 clusters from top to bottom.}
\label{SC_default} \end{figure}

We begin by estimating the extent to which conflated Web of Knowledge Subject Categories accurately describe the content of astrobiology publications.  In Figure~\ref{SC_default} we visualize the results of clustering the abstract data before sampling as described in Section~\ref{method_sc}.  The same data is presented numerically in Table~5 in the online supplement.  If SCs accurately reflect shared topical content of documents assigned to them, when the abstracts are clustered we should expect each SC to be primarily assigned a single cluster.  However, when abstracts are assigned one of five clusters (Figure~\ref{SC_default}-top panel), we observe that the cluster membership for most SCs is heterogeneous: there is no clear correspondence between a cluster and a single dominant SC.  Even the most common SC, Astronomy \& Astrophysics, is primarily distributed across the first three clusters, but is represented in all five.

Table~5 does suggest some areas in which SCs may be more appropriate document labels.  For example, Oceanography appears in only one cluster, and the Multidisciplinary Sciences SC is fairly evenly distributed across four of the five.  However, when increasing the number of clusters to 10, 15, and 20 (Figure~\ref{SC_default}), the heterogeneity of SCs within an individual cluster becomes even more pronounced. 

\begin{figure}[tbp] \begin{center}
\includegraphics[width=0.5\textwidth]{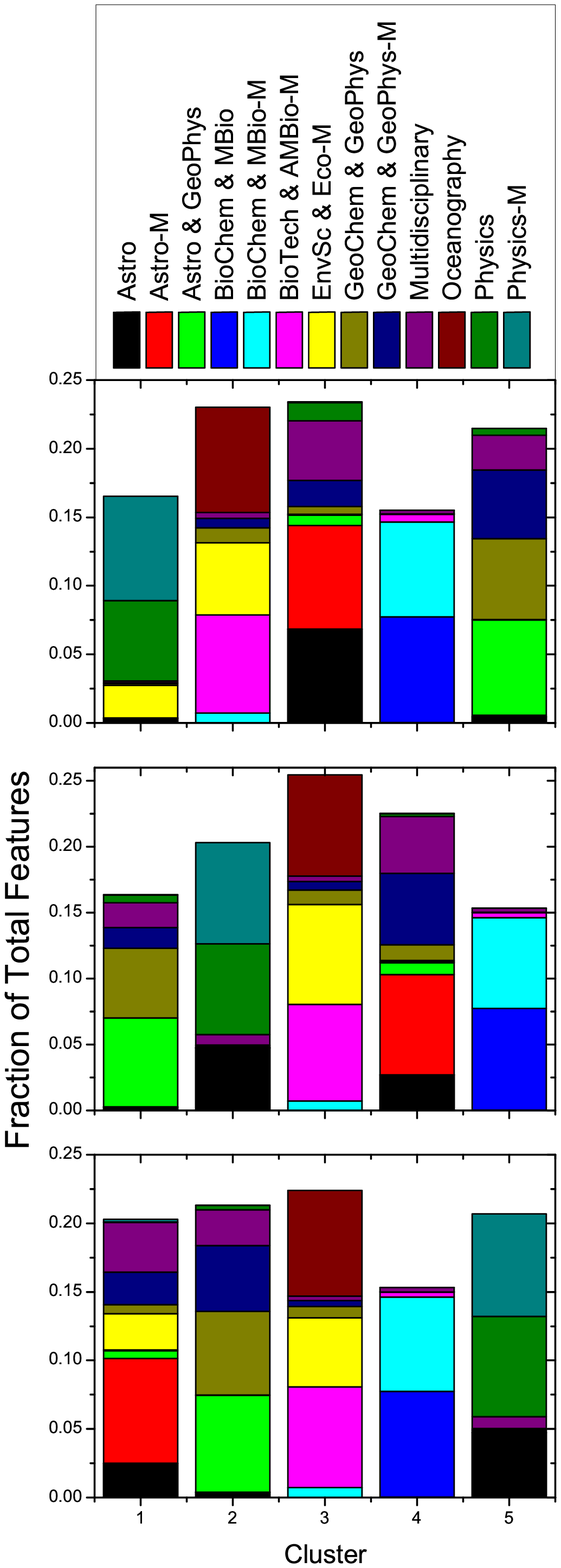} \end{center}
\caption{The results of clustering the \emph{conflated\_SC\_sampled} dataset in three separate trials.  Each abstract is assigned one of 5 clusters.}
\label{SC_05c-all_trials} \end{figure}

The dominance of the Astronomy \& Astrophysics SC in the \emph{conflated\_SC\_default} dataset suggests that we also examine the cluster relationships after the dataset has been sampled.  Figure~\ref{SC_05c-all_trials} shows the distribution of SCs in 5 clusters over 3 trials, where the cluster results of each trial are not related to each other.  For example, in successive trials, the same abstract may be assigned to different clusters; for the Astrophysics \& Geophysics SC, each trial results in different cluster assignments, though the overall distribution of clusters is roughly equal, suggesting that there is little variability between trials.  

We present the data in Figure~\ref{SC_05c-all_trials} as a fraction of the total number of features in the \emph{conflated\_SC\_sampled} dataset, where the number of features representing each SC are equal  (Figure~\ref{real_synthetic_data}).  Furthermore, the data in Table~6 in the online supplement  displays the data somewhat differently, where the fraction of features in each SC that are found in a given cluster is presented.  Examining the data with the tables provided in the online supplement makes interpreting the results easier in some instances.  In Figure~\ref{SC_05c-all_trials} and Table~6, the Astronomy, Biochemistry \& Microbiology, and Physics SCs consistently cluster with their Multidisciplinary counterparts.  Therefore, on the 5 cluster level, SCs seem to reasonably classify individual publications.     

\begin{figure}[tbp] \begin{center}
\includegraphics[width=0.5\textwidth]{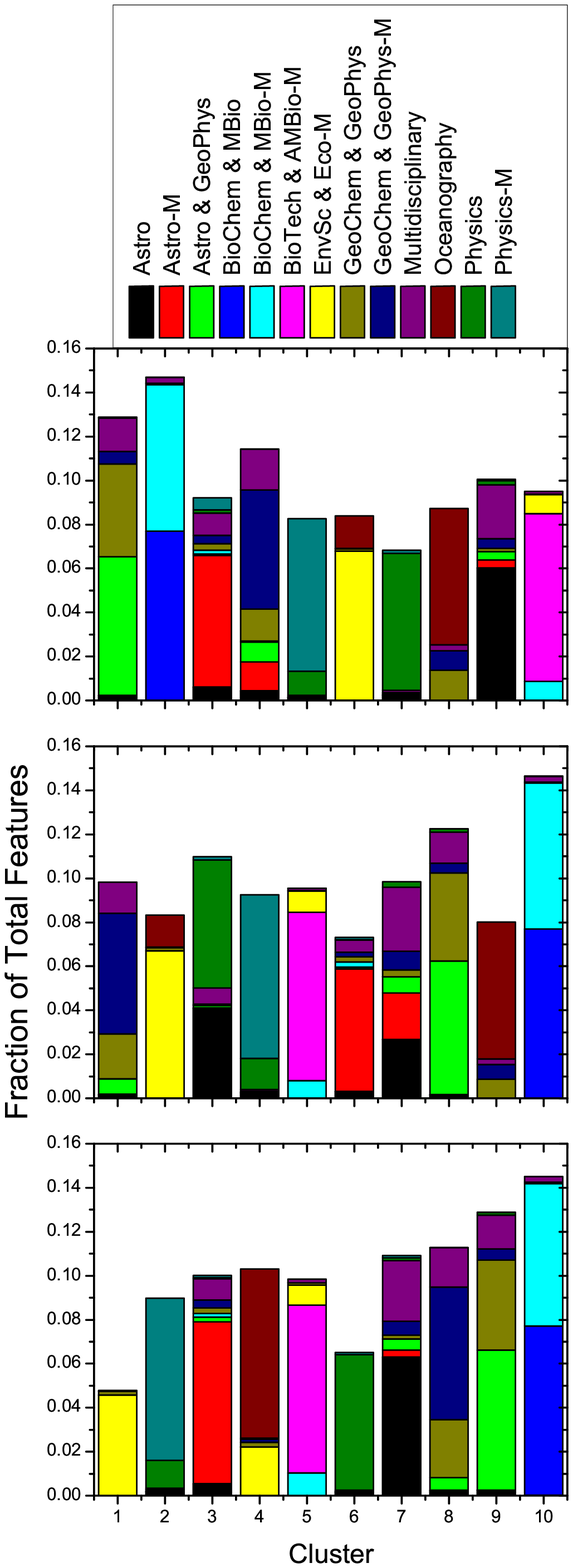} \end{center}
\caption{The results of clustering the \emph{conflated\_SC\_sampled} dataset in three separate trials.  Each abstract is assigned one of 10 clusters.}
\label{SC_10c-all_trials} \end{figure}

Since five clusters may not be sufficient to reflect the diversity of content within astrobiology, we increase the number of clusters in subsequent trials.  One would intuitively expect more SC heterogeneity within each cluster; however, increasing the number of clusters also allows more potential of each SC to dominate a single cluster. When we increase the number of clusters to 10 (Figure~\ref{SC_10c-all_trials}, Table~7), we find that most of the SCs disperse into multiple clusters.  One way to interpret this result is that more clusters allow finer distinctions between content to be revealed.  For example, Physics and Physics-Multidisciplinary, which cluster together in each trial at the 5 cluster level, tend to cluster separately at the 10 cluster level. However, the Biochemistry \& Microbiology SC and its multidisciplinary variant continue to have their abstracts cluster together.  Moreover, from Figure~\ref{SC_10c-all_trials}, we observe that the abstracts in the Biotechnology \& Applied Microbiology-Multidisciplinary SC consistently cluster together.  At the 10 cluster level, more clusters contain single dominant SCs than at the 5 cluster level.

\begin{figure}[tbp] \begin{center}
\includegraphics[width=0.5\textwidth]{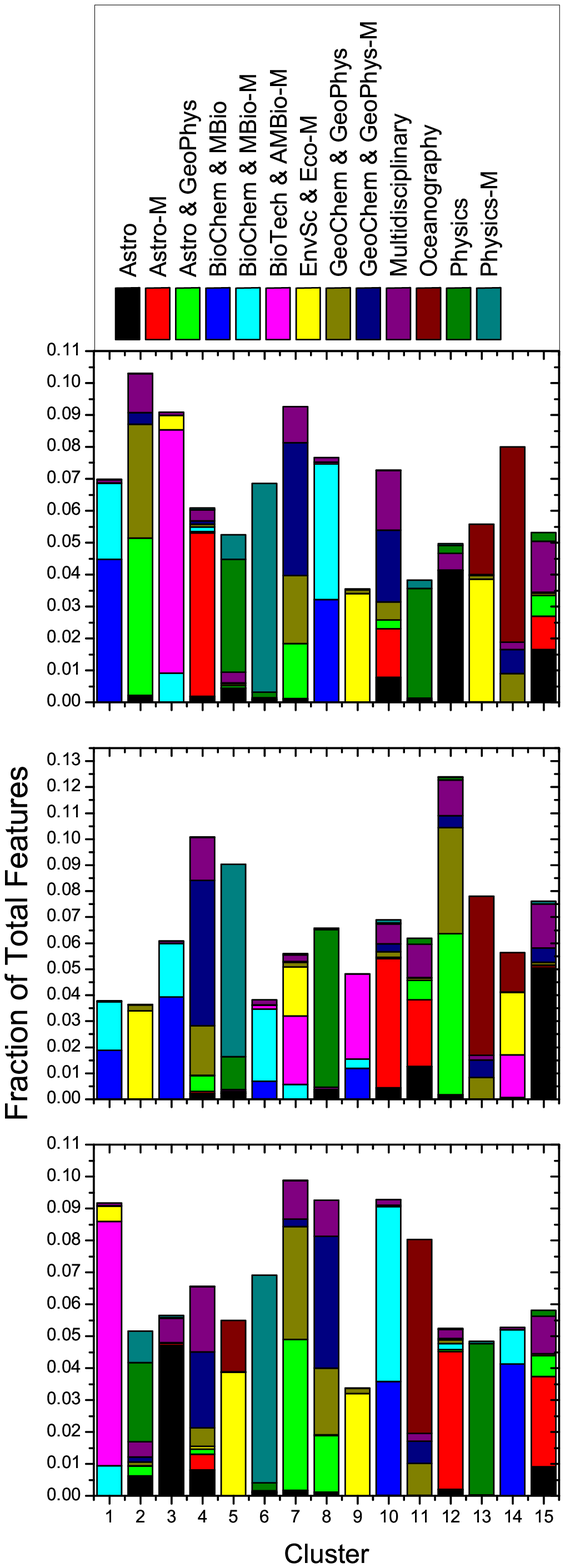} \end{center}
\caption{The results of clustering the \emph{conflated\_SC\_sampled} dataset in three separate trials.  Each abstract is assigned one of 15 clusters.}
\label{SC_15c-all_trials} \end{figure}

\begin{figure}[tbp] \begin{center}
\includegraphics[width=0.5\textwidth]{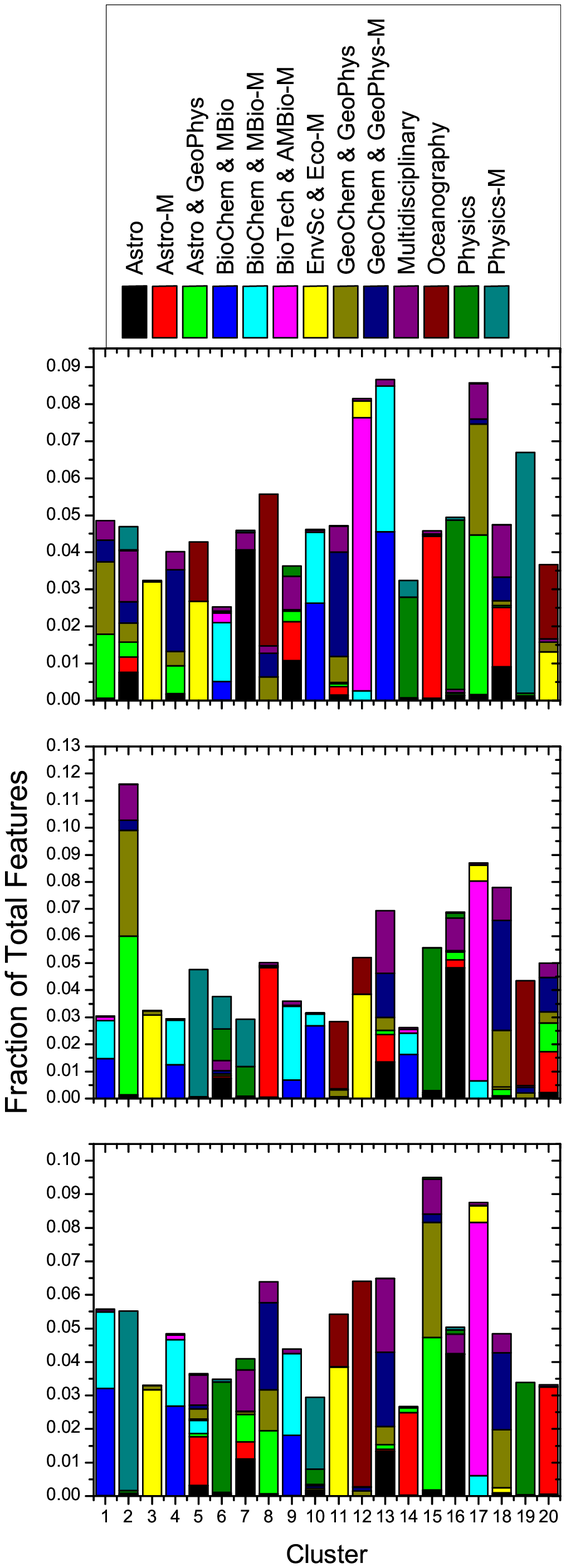} \end{center}
\caption{The results of clustering the \emph{conflated\_SC\_sampled} dataset in three separate trials.  Each abstract is assigned one of 20 clusters.}
\label{SC_20c-all_trials} \end{figure}

Figure~\ref{SC_15c-all_trials} (Table~8) and Figure~\ref{SC_20c-all_trials} (Table~9) present the results of clustering the abstracts into 15 and 20 clusters, respectively.  We observe that many of the SCs are found distributed in multiple clusters.  For example, at the 20 cluster level, what had been homogeneous cluster membership in the Biochemistry SCs at the 10 cluster level is split into three or more clusters, neither of which is shared across any other SC. 
Therefore, at these clustering levels, we operationalize a dominant SC within a cluster as one that either constitutes 50\% or more of the abstracts alone, or one that is within 50\% of the size of the most common SC\footnote{For example, a cluster with SCs constituting 30\%, 18\%, 16\% and 12\% of the abstracts would have three dominant SCs.}. By this approximation, the results at the 10 cluster level hold: as a group, the Biochemistry and Biotechnology-related SCs dominate the fewest clusters; the Astronomy, Oceanography and Physics group slightly more, and the Geochemistry and Geophysics SCs are again the most diverse, short of the Multidisciplinary Sciences SC. Overall, at the 10 cluster level, more clusters contain single dominant SCs than at the 5, 15 or 20 cluster levels, and the usefulness of SCs as document labels reaches a relative maximum.

In some cases, the trial processes reveal some inconsistencies in the cluster membership of SCs. For example, in the Biotechnology \& Applied Microbiology-Multidisciplinary SC, one would expect to have diverse membership at the 15 cluster level (Figure~\ref{SC_15c-all_trials}). However, the Biotechnology \& Applied Microbiology-Multidisciplinary SC is dominant in one cluster in trials 1 and 3, and is dominant in three clusters in trial 2.   While these results may be an artifact of the sampling and multiple-trials processes, we would expect and find that the two related SCs, Biochemistry \& Molecular Biology and Biochemistry \& Molecular Biology-Multidisciplinary are found mostly within the same clusters. This observation also holds for the Geochemistry \& Geophysics and Geochemistry \& Geophysics-Multidisciplinary SCs. The multidisciplinary SC variants (BioChem \& MBio, BioChem \& MBio-M and GeoChem \& GeoPhys, GeoChem \& GeoPhys-M) are slightly more diverse than their associated core SC, but there is a high degree of similarity between the abstracts in these two sets of related SCs.  Therefore, we conclude that even with some observed inconsistencies, the clusterer is collocating related abstracts across related SCs.

Certain related SCs tend to consistently cluster together, which suggests that SCs are sufficient for characterizing astrobiology publications.  However, other SCs have a limited effectiveness as document labels in this interdisciplinary domain, as some SCs did not map well to successively smaller cluster sizes.  Therefore, our results suggest that WoK SCs may not consistently reflect the diverse content of astrobiology publications. 

\subsection{Utilizing Subject Categories to Assess Interdisciplinarity}
In this section, we attempt to leverage the heterogeneity of SC cluster membership to assess the interdisciplinarity of astrobiology publications, and analyze only the sampled data to de-emphasize the dominance of the Astronomy \& Astrophysics SC in the dataset.  Furthermore, many observations are similar to those discussed in Section~\ref{results_SC_DL}; therefore, we will only mention in brief possible interdisciplinary connections that can be found by utilizing SCs.

In Figure~\ref{SC_05c-all_trials}, the clustering technique correctly collocates obviously similar SCs across all trials, but also identifies some less obvious potential interdisciplinary connections.  For example, across all trials in Table~6, 9\% of the abstract data in Biochemistry \& Molecular Biology--Multidisciplinary clusters with 11-14\% of the abstracts in Geochemistry \& Geophysics.  Given a document corpus of relatively equal SC distribution, as we have approximated here by the sampling process, these results suggest that papers from different SCs that consistently cluster together should be targeted for investigation by researchers interested in potential connections between the two fields. 

Across all three trials at the 10 cluster level in Figure~\ref{SC_10c-all_trials}, a single clearly dominant SC could be identified in 27 of the 30 clusters.  The Astronomy, Oceanography and Physics SCs demonstrated somewhat less monodisciplinary dominance at the 10 cluster level; all had roughly 20\% of their abstracts assigned to other clusters. The Geochemistry \& Geophysics and Environmental Sciences SCs demonstrated the most diversity apart from the pure Multidisciplinary Sciences SC, though somewhat surprisingly, the Geochemistry \& Geophysics-Multidisciplinary SC appeared in fewer clusters than its core SC.

These findings yield several possible interpretations and applications.  We would expect all astrobiology researchers to publish and cite primarily within their home disciplines, but as these results suggest, the norms of disciplinary diversity vary by field.  A potential application of this approach is a field-specific baseline metric of interdisciplinarity, a method by which an individual's research output can be compared to others in the same field in terms of the potential interdisciplinary applicability of their work.  This process could also result in an aggregate metric of interdisciplinarity for research teams via their past published work, while addressing the primary goal of discovering latent connections between the work of diverse researchers for the present and future.

Analyzing the heterogeneous cluster membership of publications from diverse SCs is one way to assess interdisciplinary research possibilities, but the probabilistic nature of this method should be emphasized.  A heterogeneous cluster could indicate that SCs are poor document labels, or that the clustering level should be adjusted to better match the data and metadata, or that a potential interdisciplinary relationship exists.  In either case, this process could inform targeted, iterative investigation. 

\subsection{Text Mining the Aggregated Abstracts} 
\begin{figure}[tbp] \begin{center}
\includegraphics[width=1\textwidth]{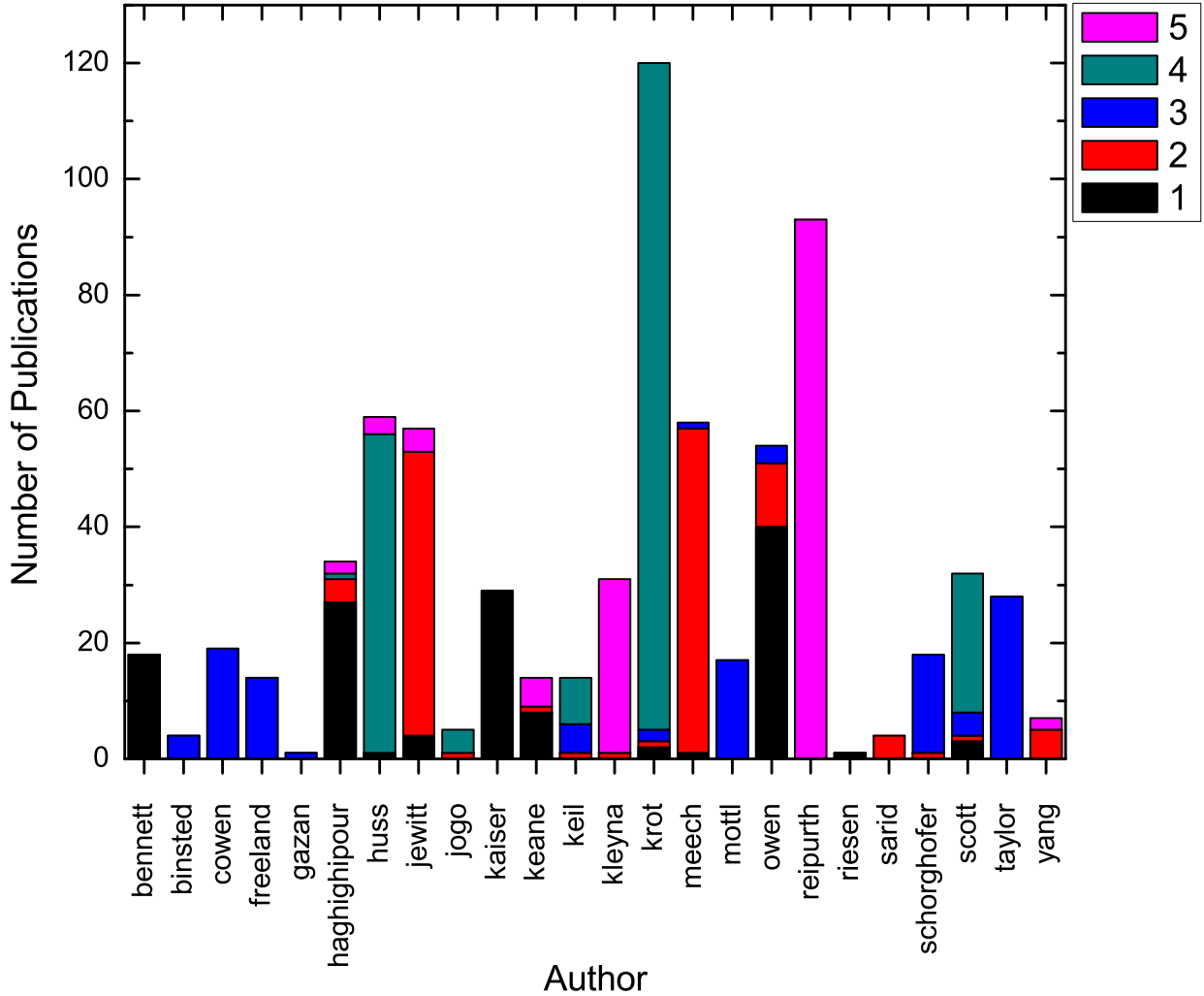} \end{center}
\caption{Clustering the \emph{aggregated\_abstracts} dataset using 5 clusters.  This plot
ensures that we are not obtaining extremely unlikely correlations and shows
that researchers from the same academic department are largely
clustering together.  For example, Bennett is a post-doctoral fellow working with Kaiser;
they often publish together and their aggregated abstracts are
clustering entirely in cluster 1.  As another example, the
geologists/geophysists Krot, Keil, Huss, Scott and Jogo are all strongly
represented in cluster 4.  The one exception is Taylor, who appears to
be clustering more strongly with the two oceanographers (Cowen and
Mottl).   As expected, researchers have the most in common with those in
their home discipline.} \label{aa_5_clusters} \end{figure}

The sIB technique was
employed to cluster the abstracts of the publications by the UHNAI team
and the references within these publications.
Figure~\ref{aa_5_clusters}, shows the results of clustering the data
into 5 clusters.   The results indicate that authors from their
respective home disciplines cluster together (see Table~\ref{tab:authors} for the list of authors and their respective home disciplines).  For example, the
geologists Krot, Keil, Huss, Scott, and Jogo are strongly represented in
cluster 4.  One exception is Taylor (geologist) who clusters with the
oceanographers (Cowen and Mottl).  Additionally, Sch\"{o}rghofer (an
astronomer by departmental affiliation) also clusters with the
oceanographers.  Furthermore, the astrochemists (Bennett and Kaiser)
have all of their publications in cluster 1.  This result suggests that
the sIB technique is able to cluster similar research on a high-level;
however, utilizing more clusters should provide a lower-level view of
overlap in research interests between the authors.

\begin{figure}[tbp] \begin{center}
\includegraphics[width=1\textwidth]{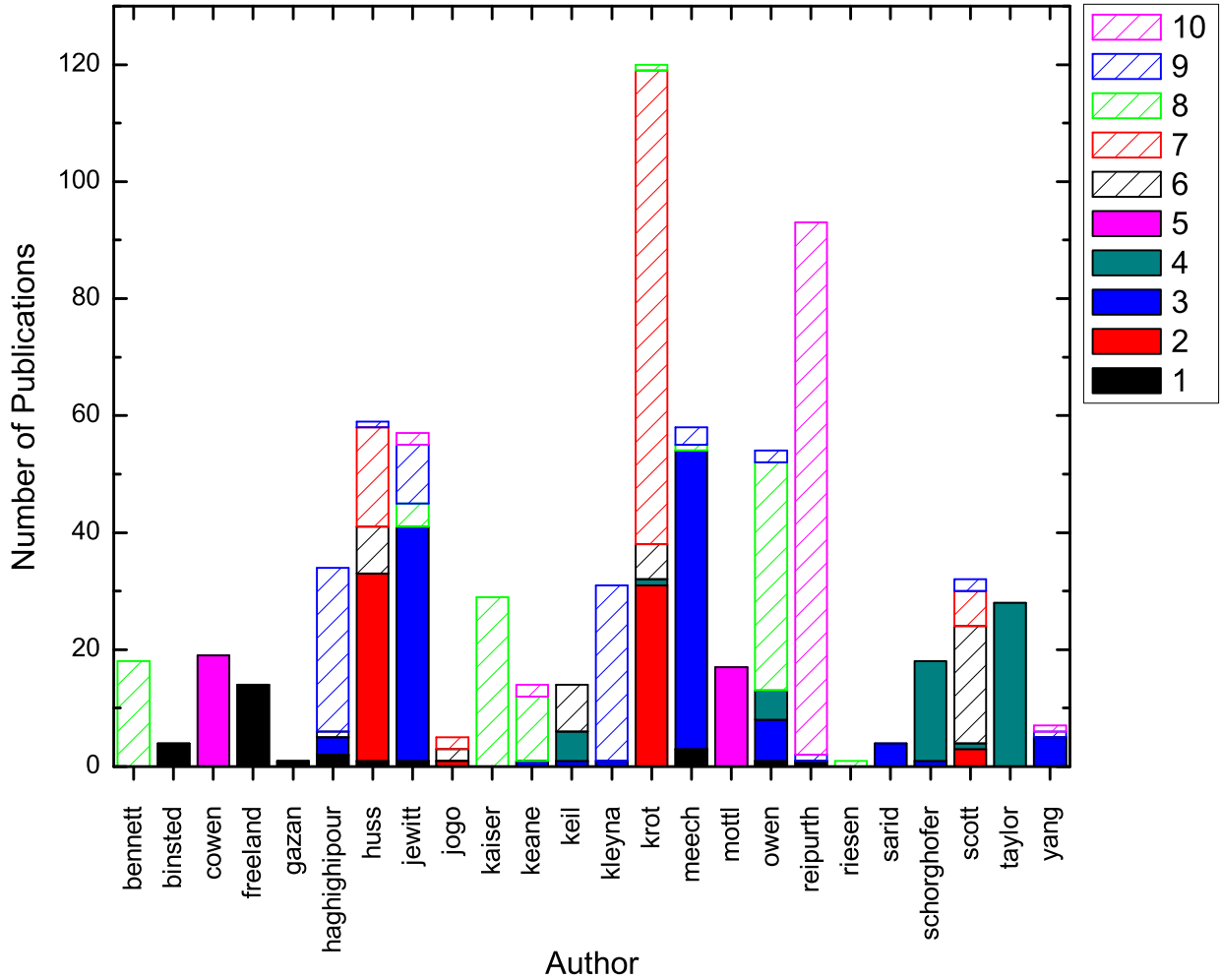} \end{center}
\caption{Clustering the \emph{aggregated\_abstracts} dataset using 10 clusters. The two oceanographers (Cowen and
Mottl) have all of their papers clustering together in Cluster 5.  The
same is true of Bennett and Kaiser (Astrochemistry).  In the previous
figure, Taylor was clustering with the oceanographers.  However, we can
see here that Taylor's work is similar to that of Sch\"{o}rghofer's, despite their membership to different home disciplines (Geology, and Astronomy respectively). Rather striking is the mono-disciplinarity regarding Reipurth's research.}
\label{aa_10_clusters} \end{figure}

When running the sIB technique for 10 clusters, we begin to see where
researchers may find potential collaboration opportunities, and we
observe which authors have specialized or broad research interests.
Research can be specialized but still integrate methods, techniques and
data from multiple disciplines. We believe that an author who is
represented primarily in a single cluster may not be engaging frequently
in interdisciplinary research, or may be focusing on narrow research
problems, or using similar research methods or equipment.  In
Figure~\ref{aa_10_clusters}, we see that the two astrochemists (Bennett
and Kaiser) are entirely represented by cluster 8, consistent with the
results presented in Figure~\ref{aa_5_clusters}. We know that their
research is heavily influenced by their experimental apparati, thus
suggesting that the experimental methods and apparati significantly
affect the description of a research track. Interestingly, Sch\"{o}rghofer's
research is on various planetary bodies such as Mars and the Moon, which
is also true of Taylor.  Therefore, clustering the text of the
aggregated abstracts sufficiently illuminates similarities in research
tracks across disciplinary boundaries, in this case, between astronomy
and geology.

\begin{figure}[tbp] \begin{center}
\includegraphics[width=1\textwidth]{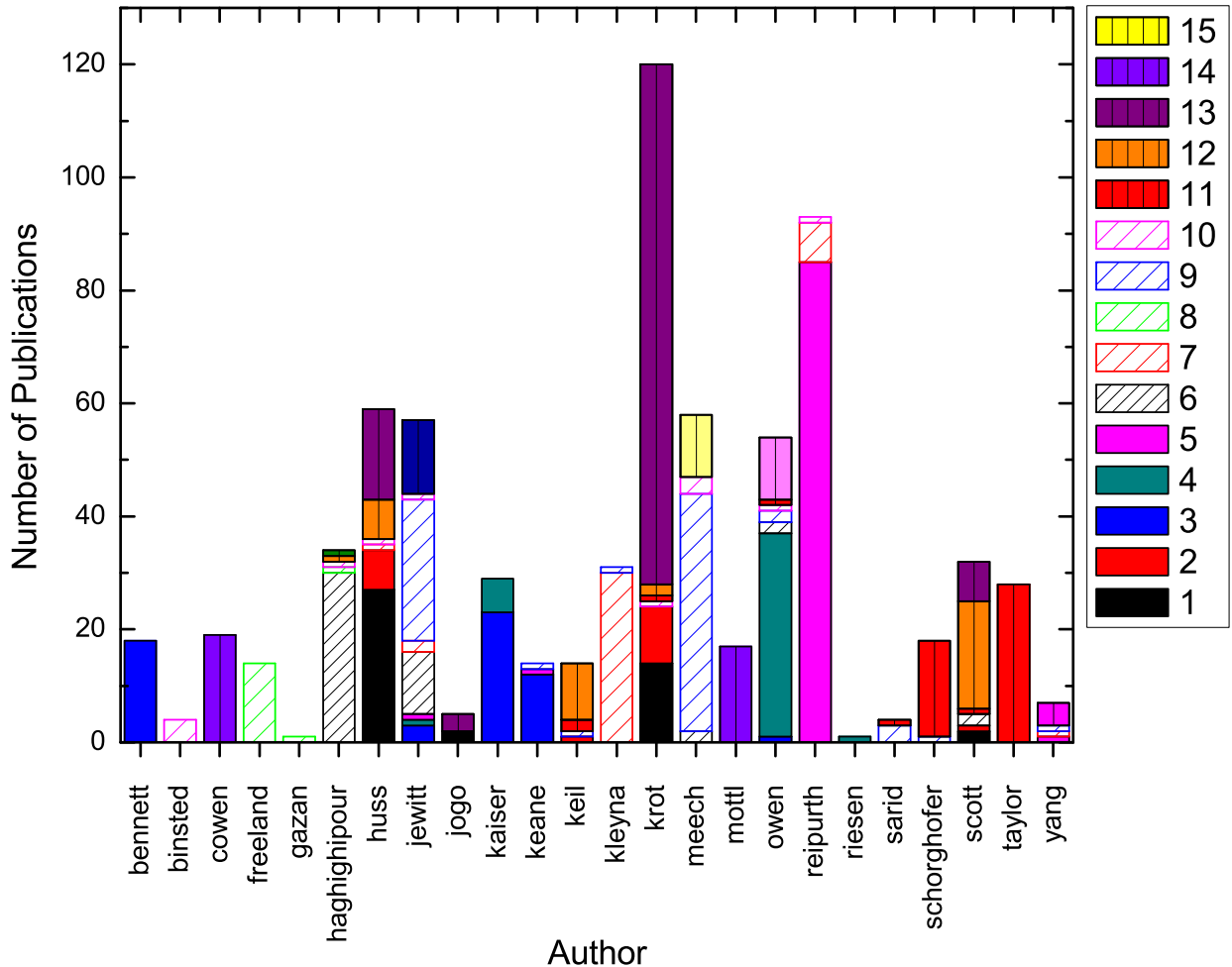} \end{center}
\caption{Clustering the \emph{aggregated\_abstracts} dataset using 15 clusters.  In this figure, Bennett and Kaiser
are no longer entirely represented by a single cluster.  When we utilized
5 and 10 clusters, Binsted and Gazan (Computer Science) and Freeland
(Biology) had their publications cluster together.  We know in
particular that the research by the computer scientists is likely to be the most dissimilar to all authors from other home disciplines.  However, when
clustering with 15 clusters, we observe Binsted's research depart from the
cluster that contains Gazan and Freeland's research and that the research has a tangential
relation to research produced by other team members.} \label{aa_15_clusters} \end{figure}

\begin{figure}[tbp] \begin{center}
\includegraphics[width=1\textwidth]{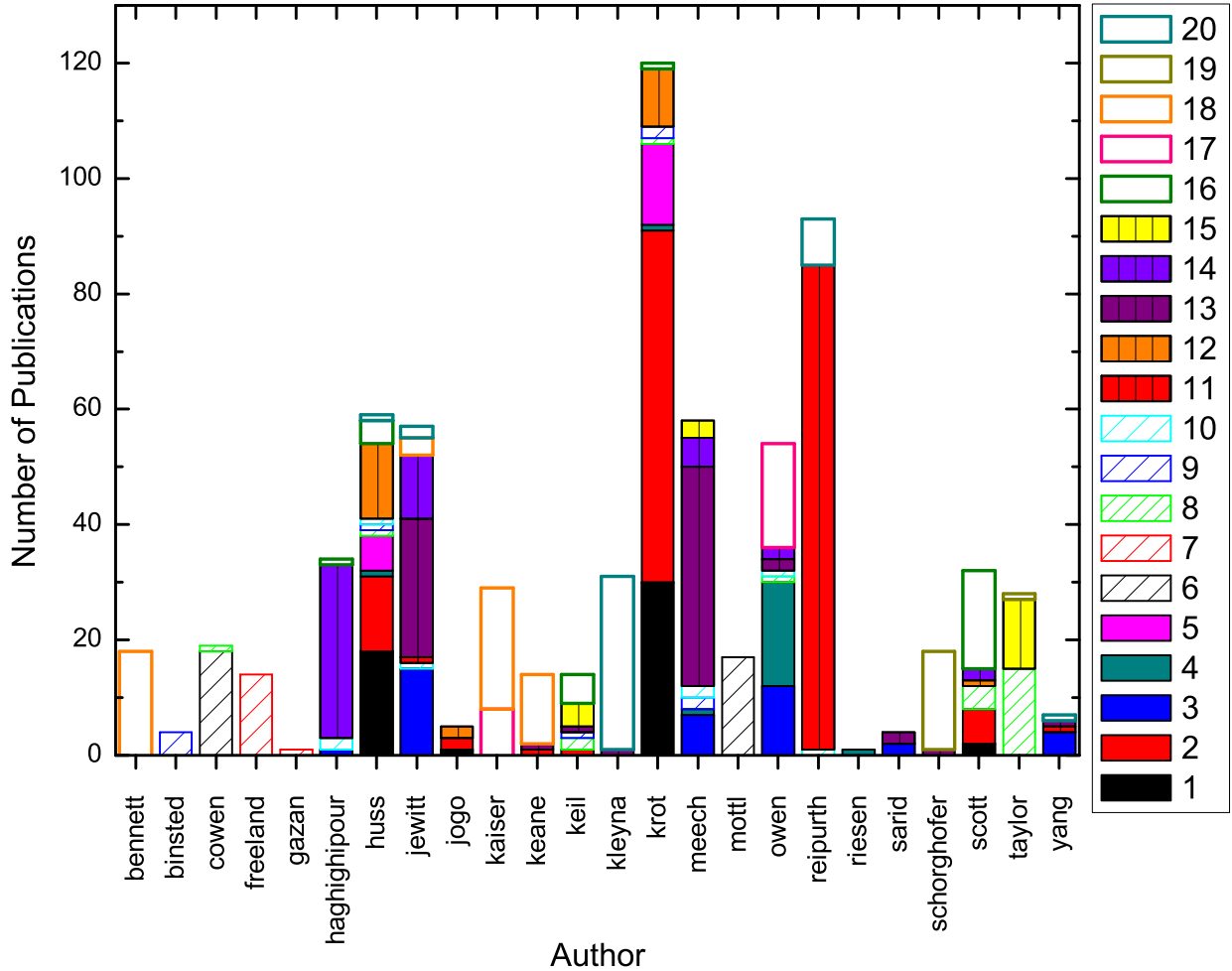} \end{center}
\caption{Clustering the \emph{aggregated\_abstracts} dataset using 20 clusters. If we assume that membership to many clusters indicates a high degree of interdisciplinarity, Huss is the most interdisciplinary UHNAI
team member.  Of the senior Astronomers (Reipurth, Meech, Jewitt,
Haghighipour, Owen, Sch\"{o}rghofer) half (Meech, Jewitt, and Owen) are
fairly diverse in their research interests, or engage in IDR, and the other half (Reipurth,
Haghighipour, Sch\"{o}rghofer) are engaged in specialized or
mono-disciplinary research.  As younger researchers, the UHNAI post-doctoral fellows
appear to be engaging in interdisciplinary research.}
\label{aa_20_clusters} \end{figure}

In Figure~\ref{aa_15_clusters}, we observe that Huss, Jewitt, Krot and
Meech's research is found in many clusters.  This signifies that their
research is likely to be very interdisciplinary.  With regards to those
authors represented by a few clusters, we cannot conclude that their
research is absolutely mono-disciplinary, as it may be very specialized,
or utilize the same methods or apparati.  However, we believe that those
UHNAI authors with publications in multiple clusters are \emph{more
likely} to be engaged in interdisciplinary research.  In
Figure~\ref{aa_20_clusters}, we observe that of the senior
(non-postdoctoral fellows) astronomers (Reipurth, Meech, Jewitt,
Haghighipour, Owen, Sch\"{o}rghofer) half (Meech, Jewitt, and Owen) are
fairly diverse in their research interests and the other half (Reipurth,
Haghighipour, Sch\"{o}rghofer) are engaged in specialized or
mono-disciplinary research.

These results suggest that the sIB method, in combination with
aggregated abstracts, can illuminate areas of implicit commonality where the research areas of scientists from diverse disciplines overlap.  Furthermore, while clusters do not inherently relate any information
about a researcher's discipline, it is clear that researchers from the
same department often cluster together.  Therefore, we expect that
performing a similar analysis on the entire NASA Astrobiology Institute
will show where collaborations between researchers can occur, and can
assist NASA with outlining research priorities.  These results can
serve as the framework for a geospatial visualization of common yet
unconnected research tracks and potential collaborators, similar to the
``hot regions'' described by \citet{Bornmann2011547}.

\section{Discussion and Conclusions} 
We clustered astrobiology abstract
data to evaluate SCs as document labels.  We attempt to
reconcile clustering (bottom-up approach) with pre-defined categories (top-down approach).
The clusters produced by text mining the abstract data did not generally
correspond well to the SCs.  Therefore, we conclude that SCs are not
well suited to the classification of astrobiology publications, and speculate that
this may also be true for other interdisciplinary fields.  One explanation is that astrobiology research outputs cite mono-disciplinary and interdisciplinary publications which may prevent SCs from forming cohesive clusters. Additionally, as
discussed in \citet{springerlink:10.1007-s11192-009-0121-z}, many
journals publish highly diverse content, which no journal-level classification system
could represent completely.  The class imbalance problem in our dataset requires us to explore utilizing an oversampling technique.  While we believe that the method remedies the skewed distribution of conflated SCs in our dataset, performing a text mining clustering analysis on a balanced astrobiology dataset without oversampling may produce different results.  That is, SCs may be more accurate when the distribution of SCs is uniform without oversampling using synthetic data.  Nonetheless, the distribution of departmental affiliations of the UHNAI researchers is skewed, which affects the distribution of publications across different SCs; it is likely that this scenario will be consistent with the other NASA Astrobiology Institute teams.   

Our results suggest that 10 clusters may be the most appropriate level at which to analyze the astrobiology collection (Figure~\ref{SC_10c-all_trials}).  Too few clusters and the interdisciplinary diversity of the source documents is not well represented; too many and they may be oversegregated, lessening the chance to identify potential commonalities in documents from different disciplines and SCs.  We suggest that when documents from different SCs cluster together, this may indicate implicit interdisciplinary connection, where knowledge in one field might inform another.  Having researchers from the constituent disciplines evaluate these common documents may provide one mechanism by which interdisciplinary science can take place, and provide a starting point for potentially productive interdisciplinary collaborations.

Similarly, text mining the aggregated abstracts using the sIB method is also suited to
the task of finding collaboration opportunities. Our experiments
consistently showed that authors from the same academic department
tended to have their publications cluster together.  If this were not
the case, we would be unable to make any claims regarding the similarity
of publications within a given cluster.  We suggest that authors whose publications cluster together could collaborate productively. An author that has publications in many clusters indicates that they are engaged
in interdisciplinary research, or perhaps that they are not, but should be.  Those authors with few publications were
either underrepresented in WoK and ADS, or were post-doctoral fellows at
the UHNAI.  We find that the majority of publications by UHNAI investigators and post-doctoral fellows appear in multiple clusters, providing evidence of actual or potential interdisciplinary research.   This is an encouraging result, as promoting boundary-crossing scientific research is one of the goals of the NASA Astrobiology Institute. Younger generations of researchers will need to synthesize techniques
from multiple disciplines to answer some of the most fundamental
questions in science in general, and astrobiology in particular.

We insinuated that a strong conclusion cannot be made regarding those
authors that are strongly represented in a single cluster.  Research in
this context is either: 1) interdisciplinary but specialized, perhaps
incorporating a synthesis between methods, techniques and data from
multiple disciplines, but with a narrow scope or 2) mono-disciplinary.
Distinguishing between these two cases requires studying the individual
works in each cluster.  Additionally, such an analysis would lead to
narrowing the scope of collaboration between two or more researchers
that are found within a single cluster.  This analysis will be explored
in future work.

The context of the interdisciplinary field of astrobiology has permitted
us to explore a method of measuring interdisciplinarity, and
identify potential collaboration opportunities.  We find that most of the UHNAI team are engaged in interdisciplinary research, and that our method suggests where productive interdisciplinary collaborations could occur.  We believe our method, which combines bibliometrics and machine learning, makes valid predictions, based on our a priori knowledge of the structure of the research team and those intra-team collaborations that currently exist.  Bibliometric studies of interdisciplinarity can benefit when augmented with machine learning algorithms, in an attempt to understand the fine-grained details of interdisciplinary research.

\section*{Acknowledgements}
We thank David Schanzenbach for devising scripts, and Mahdi Belcaid and the anonymous reviewers for insightful comments.  This material is based upon work supported by the National Aeronautics and Space Administration through the NASA Astrobiology Institute under Cooperative Agreement No. NNA08DA77A issued through the Office of Space Science.

\bibliographystyle{plainnat}

\newpage

\newpage
\pagestyle{plain}

\clearpage
\begin{spacing}{1.0}
\begin{longtable}{p{8cm}p{8cm}}
\caption{Mapping of WoK subject categories to conflated subject categories.}\\

\hline\hline
\textbf{Conflated Subject Category}              & \textbf{WoK Source Subject Category} \\
\hline 
\endfirsthead

\hline\hline
\textbf{Conflated Subject Category}              & \textbf{WoK Source Subject Category} \\
\hline 
\endhead

Agricultural Engineering-Multidisciplinary&Agricultural Engineering \textbf{with} Fisheries\\
\hline
Astronomy \& Astrophysics&Astronomy \& Astrophysics\\
\hline
Astronomy \& Astrophysics-Multidisciplinary & Astronomy \& Astrophysics-Multidisciplinary \textbf{or} Astronomy \& Astrophysics \textbf{with}: Mechanics; Engineering-Aerospace; History; Multidisciplinary Sciences; Physics; Spectroscopy\\
\hline
Astrophysics \& Geophysics &	Astronomy \& Astrophysics \textbf{with}: Geochemistry \& Geophysics; Geosciences-Multidisciplinary\\
\hline
Biochemistry \& Molecular Biology &Biochemistry \& Molecular Biology\\
\hline
Biochemistry \& Molecular Biology-Multidisciplinary&	Biochemistry \& Molecular Biology \textbf{with}: Biochemical Research Methods; Chemistry, Analytical; Biotechnology \& Applied Microbiology; Mathematical \& Computational Biology; Biology; Biophysics; Cell Biology; Computer Science, Interdisciplinary Applications; Genetics \& Heredity; Medicine, Research \& Experimental; Chemistry, Analytical; Chemistry, Medicinal; Chemistry, Organic; Pharmacology \& Pharmacy; Evolutionary Biology; Microbiology; Immunology; Infectious Diseases\\
\hline
Biology	&Biology\\
\hline
Biology-Multidisciplinary	&Biology \textbf{with}: Ecology; Evolutionary Biology; Environmental Sciences; Mathematical \& Computational Biology\\
\hline
Biology \& Geology & Biology \textbf{or} Environmental Sciences \textbf{with} Geosciences, Multidisciplinary\\
\hline
Biotechnology \& Applied Microbiology&	Biotechnology \& Applied Microbiology\\
\hline
Biotechnology \& Applied Microbiology-Multidisciplinary&	Biotechnology \& Applied Microbiology \textbf{with}: Food Science \& Technology; Microbiology; Genetics \& Heredity; Marine \& Freshwater Biology\\
\hline
Chemistry & Chemistry; Chemistry, Analytical; Chemistry, Physical; Chemistry, Organic; Chemistry, Inorganic \& Nuclear\\
\hline
Chemistry-Multidisciplinary &	Chemistry, Multidisciplinary \textbf{or} Chemistry \textbf{or} Chemistry, Analytical \textbf{with}: Spectroscopy; Chemistry, Medicinal; Nanoscience \& Nanotechnology; Materials Science, Multidisciplinary; Chemistry, Applied; Computer Science, Information Systems; Computer Science, Interdisciplinary Applications; Pharmacology \& Pharmacy; Environmental Sciences; Toxicology; Physics, Condensed Matter; Engineering, Chemical; Mathematics \& Computational Biology; Oceanography; Nuclear Science \& Technology; Polymer Science\\
\hline
Chemistry \& Physics	&Chemistry \textbf{with} Physics\\
\hline
Chemistry \& Physics-Multidisciplinary&	Chemistry and Physics \textbf{with} Nuclear Science \& Technology\\
\hline
Computer Science&	Computer Science\\
\hline
Computer Science-Multidisciplinary&	Computer Science, Multidisciplinary \textbf{or} Computer Science \textbf{with}: Information Science \& Library Science; Cybernetics; Computer Science, Artificial Intelligence; Computer Science, Theory \& Methods; Engineering, Electrical \& Electronic; Computer Science, Hardware \& Architecture; Computer Science, Information Systems; Computer Science, Interdisciplinary Applications; Geosciences, Multidisciplinary; Physics, Fluids \& Plasmas\\
\hline
Crystallography	&Crystallography\\
\hline
Education&	Education \textbf{or} Education \textbf{with} Multidisciplinary Sciences\\
\hline
Engineering&	Engineering; Engineering, Instruments \& Instrumentation; Engineering, Electrical \& Electronic; Engineering, Mechanical\\
\hline
Environmental Sciences \& Ecology&	Environmental Sciences \textbf{or} Ecology\\
\hline
Environmental Sciences \& Ecology-Multidisciplinary	&Environmental Sciences \textbf{or} Ecology \textbf{with}: Limnology; Evolutionary Biology; Marine \& Freshwater Biology; Microbiology; Oceanography; Engineering, Civil; Water Resources; Engineering, Environmental; Engineering, Chemical; Geology; Meteorology \& Atmospheric Sciences; Geography, Physical; Geosciences, Multidisciplinary; Soil Science; Toxicology; Fisheries\\
\hline
Genetics \& Heredity	&Genetics \& Heredity\\
\hline
Genetics \& Heredity-Multidisciplinary&	Genetics \& Heredity \textbf{with}: Ecology; Evolutionary Biology\\
\hline
Geochemistry \& Geophysics	&Geochemistry \& Geophysics\\
\hline
Geochemistry \& Geophysics-Multidisciplinary&	Geochemistry \& Geophysics-Multidisciplinary \textbf{or} Geochemistry \& Geophysics \textbf{with}: Geology; Meteorology \& Atmospheric Sciences; Mineralogy; Geography, Physical; Geosciences, Multidisciplinary; Paleontology\\
\hline
Geography	&Geography, Physical\\
\hline
Geology&	Geology\\
\hline
Geology-Multidisciplinary&	Geology \textbf{or} Geosciences, Multidisciplinary \textbf{with}: Energy \& Fuels; Engineering, Petroleum; Mineralogy; Mining \& Mineral Processing; Paleontology; Geography, Physical; Mathematics, Interdisciplinary Applications\\
\hline
Geology \& Oceanography&	Geology \textbf{with} Oceanography\\
\hline
Geophysics \& Oceanography	&Geochemistry \& Geophysics \textbf{with} Oceanography\\
\hline
Instruments \& Instrumentation	&Instruments \& Instrumentation\\
\hline
Life Sciences \& Biomedicine-Multidisciplinary&	Life Sciences \& Biomedicine, Other Topics; Multidisciplinary Sciences; Science \& Technology, Other Topics\\
\hline
Materials Science&	Materials Science\\
\hline
Materials Science-Multidisciplinary&	Materials Science, Multidisciplinary \textbf{or} Materials Science \textbf{with} Physics, Metallurgy \& Metallurgical Engineering \\
\hline
Mathematical \& Computational Biology&	Mathematical \& Computational Biology\\
\hline
Mathematics&	Mathematics; Mathematics, Applied; Statistics \& Probability\\
\hline
Medicine	&Medical Sciences; Psychology, Clinical; Medicine, General \& Internal; Public, Environmental \& Occupational Health; Sport Sciences\\
\hline
Meteorology \& Atmospheric Sciences&	Meteorology \& Atmospheric Sciences\\
\hline
Meteorology \& Oceanography&	Meteorology \& Atmospheric Sciences \textbf{with} Oceanography\\
\hline
Microbiology	&Microbiology\\
\hline
Mineralogy	&Mineralogy\\
\hline
Multidisciplinary Sciences&	Multidisciplinary Sciences\\
\hline
Neurosciences&	Neurosciences\\
\hline
Nutrition \& Dietetics&	Nutrition \& Dietetics\\
\hline
Oceanography	&Oceanography\\
\hline
Oceanography \& Marine Biology&	Oceanography \textbf{and} Marine \& Freshwater Biology \textbf{or} Limnology\\
\hline
Optics&	Optics\\
\hline
Optics-Multidisciplinary	&Optics \textbf{with}: Spectroscopy; Engineering, Multidisciplinary\\
\hline
Pharmacology \& Pharmacy&	Pharmacology \& Pharmacy\\
\hline
Physics&	Physics; Physics, Fluids \& Plasmas\\
\hline
Physics-Multidisciplinary&	Physics, Multidisciplinary \textbf{or} Physics \textbf{with}: Mechanics; Physics, Particles \& Fields; Physics, Nuclear; Nuclear Science \& Technology; Physics, Atomic, Molecular \& Chemical; Chemistry, Physical; Instruments \& Instrumentation; Optics; Thermodynamics; Energy \& Fuels\\
\hline
Psychology	&Psychology\\
\hline
Spectroscopy	&Spectroscopy\\
\hline
Spectroscopy-Multidisciplinary&	Spectroscopy \textbf{with}: Chemistry, Physical; Chemistry, Analytical; Physics, Atomic, Molecular \& Chemical\\
\hline
Virology&	Virology\\
\hline
Zoology&	Zoology\\
\hline
\label{tab:sc_mapping}
\end{longtable}
\end{spacing}

\clearpage
\begin{table} 
\caption{The distribution of conflated subject categories, their corresponding abstracts and the fraction of abstracts obtained through ADS and WoK.}
\label{tab:SCs}
\noindent\makebox[\textwidth]{
\begin{tabular}{p{10cm}p{1.8cm}p{2cm}p{1cm}p{1cm}} \hline
\hline\noalign{\smallskip}
Subject Category              & Number of abstracts&Fraction of dataset (\%) & ADS (\%) & WoK (\%) \\ 
\noalign{\smallskip}\hline\noalign{\smallskip}
Astronomy \& Astrophysics [Astro]&6914 & 67.68& 98.7&1.3\\ 
\hline
Astronomy \& Astrophysics-Multidisciplinary [Astro-M]&66 & 0.65&98.5 &1.5\\ 
\hline
Astrophysics \& Geophysics [Astro \& GeoPhys]&364 & 3.56& 93.4&6.6\\ 
\hline
Biochemistry \& Molecular Biology [BioChem \& MBio]&61 &0.6 & 0&100\\ 
\hline
Biochemistry \& Molecular Biology-Multidisciplinary [BioChem \& MBio-M]&109 &1.07 &0.9 &99.1\\
\hline
Biotechnology \& Applied Microbiology-Multidisciplinary [BioTech \& AMBio-M]&58 &  0.57& 0&100\\ 
\hline
Environmental Sciences \& Ecology-Multidisciplinary [EnvSc \& Eco-M]&66 & 0.65& 1.5&98.5\\ 
\hline
Geochemistry \& Geophysics [GeoChem \& GeoPhys]&978 & 9.6& 65.1& 34.9\\ 
\hline
Geochemistry \& Geophysics-Multidisciplinary [GeoChem \& GeoPhys-M]&491 &4.8 &45.8 &54.2\\ 
\hline
Multidisciplinary Sciences [Multidisciplinary]&830 & 8.12& 78.9&21.1\\
\hline
Oceanography &55 & 0.54& 0&100\\ 
\hline
Physics &86 & 0.84& 100&0\\ 
\hline
Physics-Multidisciplinary [Physics-M] &138 & 1.35& 98.6&1.4\\ 
\noalign{\smallskip}\hline
\end{tabular}}
 \end{table} 

\newpage
\begin{table} 
\caption{Home discipline of the authors at the University of Hawaii NASA Astrobiology Institute.  An asterisk (*) denotes a post-doctoral researcher.} 
\label{tab:authors}
\begin{tabular*}{1\textwidth}{ll} 
\hline
\hline\noalign{\smallskip}
Author              & Departmental Affiliation/Home Discipline\\ 
\noalign{\smallskip}\hline\noalign{\smallskip}
Bennett*&	Chemistry \\ 
Binsted&	Computer Science \\ 
Cowen&	Oceanography \\ 
Freeland&	 Biology\\ 
Gazan& Computer Science \\ 
Haghighipour& Astronomy	 \\ 
Huss& Geology	 \\ 
Jewitt& Astronomy	 \\ 
Jogo*& Geology	 \\ 
Kaiser& Chemistry	 \\ 
Keane*& Astronomy	 \\ 
Keil& Geology	 \\ 
Kleyna* & Astronomy\\
Krot & Geology\\
Meech & Astronomy \\
Mottl & Oceanography\\
Owen & Astronomy \\
Reipurth & Astronomy \\
Riesen* & Astronomy\\
Sarid* & Astronomy\\
Sch\"{o}rghofer & Astronomy\\
Scott & Geology \\
Taylor & Geology\\
Yang* & Astronomy\\
\noalign{\smallskip}\hline
\end{tabular*}
\end{table}

\newpage
\begin{table} 
\caption{Statistics for the UHNAI aggregated abstracts dataset.  A sample ($N=100$) of the total number of UHNAI publications is shown to estimate the completeness of the aggregated abstracts.}
\label{tab:AAs}
\begin{threeparttable}
\begin{tabular*}{1\textwidth}{llll} \hline
\hline\noalign{\smallskip}
 &ADS & WoK&Total   \\ 
\noalign{\smallskip}\hline\noalign{\smallskip}
Total UHNAI publications in the dataset&655 (89.6\%) &76 (10.4\%)& 731\\
Number of publications randomly selected & 88& 12&100\\
Total number of references found across the sample&3426 &756  &4182\\
Total referenced abstracts harvested& 2908 &386 &3294\\
Average completeness of the aggregated abstracts\tnote{1}& 77.5\%&50.3\% &74.3\%\\
\noalign{\smallskip}\hline
\end{tabular*}
\begin{tablenotes}
\item{$^1$}{The average completeness is measured as the mean of the completeness of each individual aggregated abstract in the sample.}
\end{tablenotes}
\end{threeparttable}
 \end{table}

\newgeometry{margin=0.5in}
\newpage

\captionsetup{width=\textwidth}

\begin{landscape}
\section*{Online Supplement}

\begin{spacing}{1.0}
\small

\begin{longtabu}{p{0.2cm}p{0.8cm}p{0.9cm}p{1.4cm}p{1.3cm}p{1.65cm}p{1.6cm}p{1.4cm}p{2cm}p{2cm}p{1.6cm}p{1.4cm}p{1.4cm}p{1.5cm}}
\caption{Clustering the \emph{conflated\_SC\_default} dataset with 5, 10, 15, and 20 clusters.}\\

\hline\noalign{\smallskip}
\hline\noalign{\smallskip}
&Astro&Astro-M	&Astro \& GeoPhys&BioChem \& MBio&	BioChem \& MBio-M&BioTech \& AMBio-M&EnvSc \& Eco-M&	GeoChem \& GeoPhys&GeoChem \& GeoPhys-M&	Multidisc-\newline iplinary&Oceano-\newline graphy&Physics&Physics-M\\
\hline\noalign{\smallskip}
\endfirsthead

\hline\noalign{\smallskip}
\hline\noalign{\smallskip}
&Astro&Astro-M	&Astro \& GeoPhys&BioChem \& MBio&	BioChem \& MBio-M&BioTech \& AMBio-M&EnvSc \& Eco-M&	GeoChem \& GeoPhys&GeoChem \& GeoPhys-M&	Multidisc-\newline iplinary&Oceano-\newline graphy&Physics&Physics-M\\ 
\hline\noalign{\smallskip}
\endhead
\hline\noalign{\smallskip}
\endfoot
\hline\noalign{\smallskip}
\endlastfoot


\multicolumn{2}{l}{5~Clusters}\\
\hline
1	&	2484	&	0	&	0	&	0	&	0	&	0	&	0	&	0	&	2	&	30	&	0	&	4	&	1	\\
2	&	1887	&	38	&	17	&	0	&	1	&	0	&	0	&	66	&	112	&	244	&	0	&	34	&	120	\\
3	&	2048	&	13	&	20	&	0	&	1	&	0	&	0	&	6	&	4	&	123	&	0	&	27	&	12	\\
4	&	308	&	2	&	310	&	0	&	1	&	0	&	1	&	532	&	62	&	204	&	0	&	16	&	2	\\
5	&	187	&	13	&	17	&	61	&	106	&	58	&	65	&	374	&	311	&	229	&	55	&	5	&	3	\\
\hline\noalign{\smallskip}
\multicolumn{2}{l}{10~Clusters}\\
\hline																											
1	&	558	&	17	&	6	&	0	&	0	&	0	&	1	&	68	&	116	&	163	&	1	&	2	&	3	\\
2	&	1821	&	0	&	0	&	0	&	0	&	0	&	0	&	0	&	0	&	15	&	0	&	0	&	1	\\
3	&	788	&	17	&	6	&	0	&	0	&	0	&	0	&	14	&	19	&	50	&	0	&	34	&	116	\\
4	&	626	&	10	&	12	&	0	&	0	&	0	&	1	&	3	&	2	&	79	&	0	&	6	&	0	\\
5	&	1141	&	1	&	0	&	0	&	0	&	0	&	0	&	0	&	0	&	27	&	0	&	3	&	3	\\
6	&	719	&	2	&	9	&	0	&	1	&	0	&	1	&	8	&	4	&	32	&	0	&	28	&	11	\\
7	&	168	&	2	&	274	&	0	&	0	&	0	&	0	&	492	&	39	&	166	&	0	&	11	&	1	\\
8	&	959	&	8	&	22	&	0	&	0	&	0	&	0	&	9	&	19	&	78	&	0	&	0	&	0	\\
9	&	113	&	3	&	30	&	0	&	0	&	0	&	6	&	312	&	288	&	140	&	35	&	1	&	0	\\
10	&	21	&	6	&	5	&	61	&	108	&	58	&	57	&	72	&	4	&	80	&	19	&	1	&	3	\\
\hline\noalign{\smallskip}
\multicolumn{2}{l}{15~Clusters}\\
\hline																											
1	&	46	&	2	&	222	&	0	&	0	&	0	&	0	&	408	&	30	&	70	&	0	&	3	&	0	\\
2	&	498	&	2	&	17	&	0	&	0	&	0	&	0	&	7	&	28	&	40	&	0	&	0	&	0	\\
3	&	702	&	15	&	1	&	0	&	0	&	0	&	0	&	9	&	19	&	28	&	0	&	34	&	113	\\
4	&	1093	&	0	&	0	&	0	&	0	&	0	&	0	&	0	&	0	&	9	&	0	&	0	&	0	\\
5	&	445	&	15	&	5	&	0	&	0	&	0	&	0	&	56	&	85	&	139	&	0	&	1	&	1	\\
6	&	525	&	6	&	10	&	0	&	0	&	0	&	0	&	9	&	3	&	53	&	0	&	2	&	0	\\
7	&	497	&	9	&	11	&	0	&	0	&	0	&	0	&	3	&	1	&	70	&	0	&	3	&	0	\\
8	&	276	&	5	&	56	&	0	&	1	&	0	&	1	&	94	&	17	&	131	&	0	&	12	&	5	\\
9	&	599	&	0	&	0	&	0	&	1	&	0	&	0	&	0	&	0	&	18	&	0	&	5	&	3	\\
10	&	454	&	1	&	6	&	0	&	0	&	0	&	0	&	2	&	6	&	16	&	0	&	20	&	6	\\
11	&	1017	&	0	&	0	&	0	&	0	&	0	&	0	&	0	&	0	&	14	&	0	&	0	&	2	\\
12	&	4	&	1	&	1	&	0	&	10	&	41	&	62	&	255	&	71	&	59	&	55	&	1	&	1	\\
13	&	14	&	3	&	2	&	61	&	97	&	17	&	1	&	14	&	1	&	48	&	0	&	0	&	3	\\
14	&	127	&	5	&	33	&	0	&	0	&	0	&	2	&	121	&	226	&	116	&	0	&	2	&	1	\\
15	&	617	&	2	&	0	&	0	&	0	&	0	&	0	&	0	&	4	&	19	&	0	&	3	&	3	\\
\hline\noalign{\smallskip}
\multicolumn{2}{l}{20~Clusters}\\
\hline																											
1	&	31	&	1	&	100	&	0	&	0	&	0	&	0	&	194	&	77	&	58	&	0	&	2	&	1	\\
2	&	450	&	5	&	8	&	0	&	0	&	0	&	0	&	7	&	2	&	48	&	0	&	0	&	0	\\
3	&	382	&	1	&	7	&	0	&	0	&	0	&	0	&	2	&	4	&	8	&	0	&	21	&	4	\\
4	&	861	&	0	&	0	&	0	&	0	&	0	&	0	&	0	&	0	&	9	&	0	&	0	&	0	\\
5	&	283	&	15	&	0	&	0	&	0	&	0	&	0	&	13	&	21	&	28	&	0	&	29	&	109	\\
6	&	560	&	4	&	0	&	0	&	0	&	0	&	0	&	3	&	1	&	14	&	0	&	3	&	8	\\
7	&	449	&	8	&	9	&	0	&	0	&	0	&	0	&	3	&	1	&	47	&	0	&	4	&	0	\\
8	&	713	&	0	&	0	&	0	&	0	&	0	&	0	&	0	&	0	&	15	&	0	&	2	&	1	\\
9	&	574	&	0	&	0	&	0	&	1	&	0	&	0	&	0	&	0	&	16	&	0	&	2	&	3	\\
10	&	323	&	3	&	2	&	0	&	0	&	0	&	0	&	26	&	49	&	100	&	0	&	0	&	1	\\
11	&	378	&	3	&	0	&	0	&	0	&	0	&	0	&	0	&	4	&	11	&	0	&	1	&	4	\\
12	&	7	&	2	&	1	&	61	&	103	&	53	&	22	&	24	&	0	&	59	&	2	&	0	&	2	\\
13	&	31	&	1	&	159	&	0	&	0	&	0	&	0	&	282	&	17	&	53	&	0	&	1	&	0	\\
14	&	200	&	12	&	11	&	0	&	3	&	0	&	1	&	35	&	34	&	70	&	0	&	0	&	1	\\
15	&	313	&	2	&	13	&	0	&	0	&	0	&	0	&	6	&	32	&	25	&	0	&	1	&	0	\\
16	&	222	&	1	&	40	&	0	&	0	&	0	&	0	&	79	&	3	&	93	&	0	&	15	&	3	\\
17	&	119	&	5	&	3	&	0	&	0	&	0	&	2	&	65	&	174	&	85	&	0	&	2	&	0	\\
18	&	352	&	2	&	10	&	0	&	0	&	0	&	0	&	2	&	2	&	47	&	0	&	2	&	0	\\
19	&	665	&	0	&	0	&	0	&	0	&	0	&	0	&	0	&	0	&	7	&	0	&	0	&	1	\\
20	&	1	&	1	&	1	&	0	&	2	&	5	&	41	&	237	&	70	&	37	&	53	&	1	&	0	\\
\label{tab:SC_default}
\end{longtabu}
\end{spacing}
\end{landscape}

\newpage

\begin{landscape}
\begin{spacing}{1.0}
\small

\begin{longtabu}{p{0.2cm}p{0.8cm}p{0.9cm}p{1.4cm}p{1.3cm}p{1.65cm}p{1.6cm}p{1.4cm}p{2cm}p{2cm}p{1.6cm}p{1.4cm}p{1.4cm}p{1.5cm}}
\caption{Clustering abstracts in the \emph{conflated\_SC\_sampled} dataset with 5 clusters.}\\
\hline\noalign{\smallskip}
\hline\noalign{\smallskip}

&Astro&Astro-M	&Astro \& GeoPhys&BioChem \& MBio&	BioChem \& MBio-M&BioTech \& AMBio-M&EnvSc \& Eco-M&	GeoChem \& GeoPhys&GeoChem \& GeoPhys-M&	Multidisc-\newline iplinary&Oceano-\newline graphy&Physics&Physics-M\\

\hline\noalign{\smallskip}
\endhead

\multicolumn{2}{l}{Trial~1}\\
\hline
1&0.05	&	0.00	&	0.00	&	0.00	&	0.00	&	0.00	&	0.31	&	0.02	&	0.01	&	0.02	&	0.00	&	0.76	&	0.99	\\
2&0.00	&	0.00	&	0.00	&	0.00	&	0.09	&	0.93	&	0.68	&	0.14	&	0.09	&	0.05	&	1.00	&	0.00	&	0.00	\\
3&0.89	&	0.98	&	0.10	&	0.00	&	0.01	&	0.00	&	0.00	&	0.07	&	0.25	&	0.56	&	0.00	&	0.17	&	0.01	\\
4&0.00	&	0.00	&	0.00	&	1.00	&	0.90	&	0.07	&	0.00	&	0.00	&	0.00	&	0.04	&	0.00	&	0.00	&	0.00	\\
5&0.06	&	0.01	&	0.90	&	0.00	&	0.00	&	0.00	&	0.01	&	0.77	&	0.65	&	0.33	&	0.00	&	0.07	&	0.00	\\
\noalign{\smallskip}\hline\noalign{\smallskip}
\multicolumn{2}{l}{Trial~2}\\																								
\hline
1&0.04	&	0.00	&	0.88	&	0.00	&	0.00	&	0.00	&	0.00	&	0.69	&	0.21	&	0.24	&	0.00	&	0.08	&	0.00	\\
2&0.61	&	0.01	&	0.00	&	0.00	&	0.00	&	0.00	&	0.00	&	0.01	&	0.01	&	0.10	&	0.00	&	0.90	&	1.00	\\
3&0.00	&	0.00	&	0.00	&	0.00	&	0.09	&	0.95	&	0.99	&	0.14	&	0.08	&	0.05	&	1.00	&	0.00	&	0.00	\\
4&0.35	&	0.99	&	0.12	&	0.00	&	0.01	&	0.00	&	0.01	&	0.16	&	0.70	&	0.56	&	0.00	&	0.03	&	0.00	\\
5&0.00	&	0.00	&	0.00	&	1.00	&	0.90	&	0.05	&	0.00	&	0.01	&	0.00	&	0.04	&	0.00	&	0.00	&	0.00	\\
\noalign{\smallskip}\hline\noalign{\smallskip}
\multicolumn{2}{l}{Trial~3}\\																							
\hline
1&0.33	&	0.99	&	0.07	&	0.00	&	0.01	&	0.00	&	0.34	&	0.09	&	0.31	&	0.47	&	0.00	&	0.00	&	0.03	\\
2&0.05	&	0.00	&	0.92	&	0.00	&	0.00	&	0.00	&	0.00	&	0.79	&	0.63	&	0.34	&	0.00	&	0.04	&	0.00	\\
3&0.00	&	0.00	&	0.00	&	0.00	&	0.09	&	0.96	&	0.66	&	0.11	&	0.06	&	0.05	&	1.00	&	0.00	&	0.00	\\
4&0.00	&	0.00	&	0.00	&	1.00	&	0.90	&	0.04	&	0.00	&	0.00	&	0.00	&	0.04	&	0.00	&	0.00	&	0.00	\\
5&0.62	&	0.00	&	0.01	&	0.00	&	0.00	&	0.00	&	0.00	&	0.01	&	0.01	&	0.11	&	0.00	&	0.95	&	0.97	\\

\noalign{\smallskip}\hline
\label{SC_5_clusters}
\end{longtabu}


\end{spacing}
\end{landscape}

\newpage
\begin{landscape}
\begin{spacing}{1.0}
\small

\begin{longtabu}{p{0.2cm}p{0.8cm}p{0.9cm}p{1.4cm}p{1.3cm}p{1.65cm}p{1.6cm}p{1.4cm}p{2cm}p{2cm}p{1.6cm}p{1.4cm}p{1.4cm}p{1.5cm}}
\caption{Clustering abstracts in the \emph{conflated\_SC\_sampled} dataset with 10 clusters.}\\

\hline\noalign{\smallskip}
\hline\noalign{\smallskip}
&Astro&Astro-M	&Astro \& GeoPhys&BioChem \& MBio&	BioChem \& MBio-M&BioTech \& AMBio-M&EnvSc \& Eco-M&	GeoChem \& GeoPhys&GeoChem \& GeoPhys-M&	Multidisc-\newline iplinary&Oceano-\newline graphy&Physics&Physics-M\\
\hline\noalign{\smallskip}
\endfirsthead

\hline\noalign{\smallskip}
\hline\noalign{\smallskip}
&Astro&Astro-M	&Astro \& GeoPhys&BioChem \& MBio&	BioChem \& MBio-M&BioTech \& AMBio-M&EnvSc \& Eco-M&	GeoChem \& GeoPhys&GeoChem \& GeoPhys-M&	Multidisc-\newline iplinary&Oceano-\newline graphy&Physics&Physics-M\\ 
\hline\noalign{\smallskip}
\endhead
\hline\noalign{\smallskip}
\endfoot

\multicolumn{2}{l}{Trial~1}\\
\hline
1	&	0.03	&	0.00	&	0.82	&	0.00	&	0.00	&	0.00	&	0.00	&	0.55	&	0.07	&	0.20	&	0.00	&	0.00	&	0.00	\\
2	&	0.00	&	0.00	&	0.00	&	1.00	&	0.87	&	0.01	&	0.00	&	0.00	&	0.00	&	0.03	&	0.00	&	0.00	&	0.00	\\
3	&	0.08	&	0.78	&	0.01	&	0.00	&	0.02	&	0.00	&	0.00	&	0.04	&	0.05	&	0.13	&	0.00	&	0.02	&	0.07	\\
4	&	0.06	&	0.17	&	0.12	&	0.00	&	0.00	&	0.00	&	0.01	&	0.19	&	0.70	&	0.24	&	0.00	&	0.00	&	0.00	\\
5	&	0.02	&	0.00	&	0.00	&	0.00	&	0.00	&	0.00	&	0.00	&	0.00	&	0.00	&	0.01	&	0.00	&	0.14	&	0.90	\\
6	&	0.00	&	0.00	&	0.00	&	0.00	&	0.00	&	0.00	&	0.88	&	0.02	&	0.00	&	0.00	&	0.19	&	0.00	&	0.00	\\
7	&	0.03	&	0.00	&	0.01	&	0.00	&	0.00	&	0.00	&	0.00	&	0.00	&	0.00	&	0.01	&	0.00	&	0.81	&	0.02	\\
8	&	0.00	&	0.00	&	0.00	&	0.00	&	0.00	&	0.00	&	0.00	&	0.18	&	0.12	&	0.04	&	0.81	&	0.00	&	0.00	\\
9	&	0.78	&	0.05	&	0.05	&	0.00	&	0.00	&	0.00	&	0.00	&	0.02	&	0.06	&	0.32	&	0.00	&	0.03	&	0.01	\\
10	&	0.00	&	0.00	&	0.00	&	0.00	&	0.11	&	0.99	&	0.11	&	0.00	&	0.00	&	0.02	&	0.00	&	0.00	&	0.00	\\
\noalign{\smallskip}\hline\noalign{\smallskip}
\multicolumn{2}{l}{Trial~2}\\																									
\hline
1	&	0.02	&	0.00	&	0.09	&	0.00	&	0.00	&	0.00	&	0.00	&	0.27	&	0.71	&	0.18	&	0.00	&	0.00	&	0.00	\\
2	&	0.00	&	0.00	&	0.00	&	0.00	&	0.00	&	0.00	&	0.87	&	0.02	&	0.00	&	0.00	&	0.19	&	0.00	&	0.00	\\
3	&	0.53	&	0.00	&	0.01	&	0.00	&	0.00	&	0.00	&	0.00	&	0.00	&	0.00	&	0.09	&	0.00	&	0.76	&	0.02	\\
4	&	0.03	&	0.00	&	0.00	&	0.00	&	0.00	&	0.00	&	0.00	&	0.01	&	0.00	&	0.01	&	0.00	&	0.19	&	0.97	\\
5	&	0.00	&	0.00	&	0.00	&	0.00	&	0.11	&	0.99	&	0.13	&	0.00	&	0.00	&	0.01	&	0.00	&	0.00	&	0.00	\\
6	&	0.04	&	0.72	&	0.01	&	0.00	&	0.03	&	0.00	&	0.00	&	0.03	&	0.03	&	0.07	&	0.00	&	0.00	&	0.01	\\
7	&	0.35	&	0.27	&	0.10	&	0.00	&	0.00	&	0.00	&	0.00	&	0.04	&	0.11	&	0.38	&	0.00	&	0.03	&	0.00	\\
8	&	0.02	&	0.00	&	0.79	&	0.00	&	0.00	&	0.00	&	0.00	&	0.52	&	0.06	&	0.18	&	0.00	&	0.02	&	0.00	\\
9	&	0.00	&	0.00	&	0.00	&	0.00	&	0.00	&	0.00	&	0.00	&	0.11	&	0.09	&	0.03	&	0.81	&	0.00	&	0.00	\\
10	&	0.00	&	0.00	&	0.00	&	1.00	&	0.86	&	0.01	&	0.00	&	0.00	&	0.00	&	0.03	&	0.00	&	0.00	&	0.00	\\
\noalign{\smallskip}\hline\noalign{\smallskip}
\multicolumn{2}{l}{Trial~3}\\																									
\hline
1	&	0.00	&	0.00	&	0.00	&	0.00	&	0.00	&	0.00	&	0.59	&	0.02	&	0.00	&	0.00	&	0.00	&	0.00	&	0.00	\\
2	&	0.02	&	0.00	&	0.00	&	0.00	&	0.00	&	0.00	&	0.00	&	0.01	&	0.00	&	0.01	&	0.00	&	0.16	&	0.96	\\
3	&	0.07	&	0.96	&	0.03	&	0.00	&	0.02	&	0.00	&	0.00	&	0.03	&	0.05	&	0.13	&	0.00	&	0.01	&	0.01	\\
4	&	0.00	&	0.00	&	0.00	&	0.00	&	0.00	&	0.00	&	0.29	&	0.03	&	0.02	&	0.01	&	1.00	&	0.00	&	0.00	\\
5	&	0.00	&	0.00	&	0.00	&	0.00	&	0.13	&	0.99	&	0.12	&	0.01	&	0.00	&	0.02	&	0.00	&	0.00	&	0.00	\\
6	&	0.02	&	0.00	&	0.00	&	0.00	&	0.00	&	0.00	&	0.00	&	0.00	&	0.00	&	0.01	&	0.00	&	0.80	&	0.01	\\
7	&	0.82	&	0.04	&	0.07	&	0.00	&	0.00	&	0.00	&	0.00	&	0.02	&	0.08	&	0.36	&	0.00	&	0.02	&	0.01	\\
8	&	0.03	&	0.00	&	0.08	&	0.00	&	0.00	&	0.00	&	0.00	&	0.34	&	0.78	&	0.23	&	0.00	&	0.00	&	0.00	\\
9	&	0.03	&	0.00	&	0.83	&	0.00	&	0.00	&	0.00	&	0.00	&	0.53	&	0.06	&	0.20	&	0.00	&	0.02	&	0.00	\\
10	&	0.00	&	0.00	&	0.00	&	1.00	&	0.84	&	0.01	&	0.00	&	0.00	&	0.00	&	0.03	&	0.00	&	0.00	&	0.00	\\

\label{SC_10_clusters}
\end{longtabu}
\end{spacing}
\end{landscape}

\newpage
\begin{landscape}

\begin{spacing}{1.0}
\small

\begin{longtabu}{p{0.2cm}p{0.8cm}p{0.9cm}p{1.4cm}p{1.3cm}p{1.65cm}p{1.6cm}p{1.4cm}p{2cm}p{2cm}p{1.6cm}p{1.4cm}p{1.4cm}p{1.5cm}}
\caption{Clustering abstracts in the \emph{conflated\_SC\_sampled} dataset with 15 clusters.}\\

\hline\noalign{\smallskip}
\hline\noalign{\smallskip}
&Astro&Astro-M	&Astro \& GeoPhys&BioChem \& MBio&	BioChem \& MBio-M&BioTech \& AMBio-M&EnvSc \& Eco-M&	GeoChem \& GeoPhys&GeoChem \& GeoPhys-M&	Multidisc-\newline iplinary&Oceano-\newline graphy&Physics&Physics-M\\
\hline\noalign{\smallskip}
\endfirsthead

\hline\noalign{\smallskip}
\hline\noalign{\smallskip}
&Astro&Astro-M	&Astro \& GeoPhys&BioChem \& MBio&	BioChem \& MBio-M&BioTech \& AMBio-M&EnvSc \& Eco-M&	GeoChem \& GeoPhys&GeoChem \& GeoPhys-M&	Multidisc-\newline iplinary&Oceano-\newline graphy&Physics&Physics-M\\ 
\hline\noalign{\smallskip}
\endhead
\hline\noalign{\smallskip}
\endfoot

\multicolumn{2}{l}{Trial~1}\\
\hline
1	&	0.00	&	0.00	&	0.00	&	0.58	&	0.31	&	0.00	&	0.00	&	0.00	&	0.00	&	0.01	&	0.00	&	0.00	&	0.00	\\
2	&	0.03	&	0.00	&	0.64	&	0.00	&	0.00	&	0.00	&	0.00	&	0.46	&	0.05	&	0.16	&	0.00	&	0.00	&	0.00	\\
3	&	0.00	&	0.00	&	0.00	&	0.00	&	0.12	&	0.99	&	0.06	&	0.00	&	0.00	&	0.01	&	0.00	&	0.00	&	0.00	\\
4	&	0.02	&	0.66	&	0.01	&	0.00	&	0.02	&	0.00	&	0.00	&	0.01	&	0.01	&	0.04	&	0.00	&	0.00	&	0.01	\\
5	&	0.06	&	0.00	&	0.01	&	0.00	&	0.00	&	0.00	&	0.00	&	0.01	&	0.00	&	0.04	&	0.00	&	0.46	&	0.10	\\
6	&	0.01	&	0.00	&	0.00	&	0.00	&	0.00	&	0.00	&	0.00	&	0.00	&	0.00	&	0.00	&	0.00	&	0.02	&	0.85	\\
7	&	0.01	&	0.00	&	0.22	&	0.00	&	0.00	&	0.00	&	0.00	&	0.28	&	0.54	&	0.15	&	0.00	&	0.00	&	0.00	\\
8	&	0.00	&	0.00	&	0.00	&	0.42	&	0.55	&	0.01	&	0.00	&	0.00	&	0.00	&	0.02	&	0.00	&	0.00	&	0.00	\\
9	&	0.00	&	0.00	&	0.00	&	0.00	&	0.00	&	0.00	&	0.44	&	0.02	&	0.00	&	0.00	&	0.00	&	0.00	&	0.00	\\
10	&	0.10	&	0.20	&	0.03	&	0.00	&	0.00	&	0.00	&	0.00	&	0.07	&	0.29	&	0.24	&	0.00	&	0.00	&	0.00	\\
11	&	0.01	&	0.00	&	0.00	&	0.00	&	0.00	&	0.00	&	0.00	&	0.01	&	0.00	&	0.00	&	0.00	&	0.45	&	0.03	\\
12	&	0.54	&	0.00	&	0.00	&	0.00	&	0.00	&	0.00	&	0.00	&	0.00	&	0.00	&	0.07	&	0.00	&	0.03	&	0.01	\\
13	&	0.00	&	0.00	&	0.00	&	0.00	&	0.00	&	0.00	&	0.50	&	0.02	&	0.00	&	0.00	&	0.20	&	0.00	&	0.00	\\
14	&	0.00	&	0.00	&	0.00	&	0.00	&	0.00	&	0.00	&	0.00	&	0.12	&	0.10	&	0.03	&	0.80	&	0.00	&	0.00	\\
15	&	0.21	&	0.14	&	0.08	&	0.00	&	0.00	&	0.00	&	0.00	&	0.01	&	0.00	&	0.21	&	0.00	&	0.03	&	0.00	\\
\noalign{\smallskip}\hline\noalign{\smallskip}
\multicolumn{2}{l}{Trial~2}\\
\hline																							
1	&	0.00	&	0.00	&	0.00	&	0.24	&	0.24	&	0.00	&	0.00	&	0.00	&	0.00	&	0.00	&	0.00	&	0.00	&	0.00	\\
2	&	0.00	&	0.00	&	0.00	&	0.00	&	0.00	&	0.00	&	0.44	&	0.03	&	0.00	&	0.00	&	0.00	&	0.00	&	0.00	\\
3	&	0.00	&	0.00	&	0.00	&	0.51	&	0.27	&	0.00	&	0.00	&	0.00	&	0.00	&	0.01	&	0.00	&	0.00	&	0.00	\\
4	&	0.03	&	0.01	&	0.08	&	0.00	&	0.00	&	0.00	&	0.00	&	0.25	&	0.73	&	0.22	&	0.00	&	0.00	&	0.00	\\
5	&	0.03	&	0.00	&	0.00	&	0.00	&	0.00	&	0.00	&	0.00	&	0.00	&	0.00	&	0.01	&	0.00	&	0.17	&	0.96	\\
6	&	0.00	&	0.00	&	0.00	&	0.09	&	0.36	&	0.02	&	0.00	&	0.00	&	0.00	&	0.02	&	0.00	&	0.00	&	0.00	\\
7	&	0.00	&	0.00	&	0.00	&	0.00	&	0.07	&	0.34	&	0.24	&	0.02	&	0.00	&	0.03	&	0.01	&	0.00	&	0.00	\\
8	&	0.04	&	0.00	&	0.01	&	0.00	&	0.00	&	0.00	&	0.00	&	0.00	&	0.00	&	0.01	&	0.00	&	0.79	&	0.01	\\
9	&	0.00	&	0.00	&	0.00	&	0.16	&	0.04	&	0.43	&	0.00	&	0.00	&	0.00	&	0.00	&	0.00	&	0.00	&	0.00	\\
10	&	0.06	&	0.65	&	0.01	&	0.00	&	0.00	&	0.00	&	0.00	&	0.03	&	0.04	&	0.10	&	0.00	&	0.00	&	0.02	\\
11	&	0.16	&	0.33	&	0.10	&	0.00	&	0.00	&	0.00	&	0.00	&	0.01	&	0.00	&	0.17	&	0.00	&	0.03	&	0.00	\\
12	&	0.02	&	0.00	&	0.80	&	0.00	&	0.00	&	0.00	&	0.00	&	0.53	&	0.06	&	0.18	&	0.00	&	0.01	&	0.00	\\
13	&	0.00	&	0.00	&	0.00	&	0.00	&	0.00	&	0.00	&	0.00	&	0.11	&	0.09	&	0.02	&	0.80	&	0.00	&	0.00	\\
14	&	0.00	&	0.00	&	0.00	&	0.00	&	0.01	&	0.21	&	0.31	&	0.00	&	0.00	&	0.00	&	0.20	&	0.00	&	0.00	\\
15	&	0.66	&	0.01	&	0.00	&	0.00	&	0.00	&	0.00	&	0.00	&	0.01	&	0.07	&	0.22	&	0.00	&	0.00	&	0.01	\\
\noalign{\smallskip}\hline\noalign{\smallskip}
\multicolumn{2}{l}{Trial~3}\\
\hline																											
1	&	0.00	&	0.00	&	0.00	&	0.00	&	0.12	&	0.99	&	0.06	&	0.00	&	0.00	&	0.01	&	0.00	&	0.00	&	0.00	\\
2	&	0.08	&	0.00	&	0.04	&	0.00	&	0.00	&	0.00	&	0.00	&	0.02	&	0.02	&	0.06	&	0.00	&	0.32	&	0.13	\\
3	&	0.61	&	0.01	&	0.00	&	0.00	&	0.00	&	0.00	&	0.00	&	0.00	&	0.00	&	0.10	&	0.00	&	0.00	&	0.01	\\
4	&	0.10	&	0.06	&	0.02	&	0.00	&	0.00	&	0.00	&	0.01	&	0.07	&	0.31	&	0.26	&	0.00	&	0.00	&	0.00	\\
5	&	0.00	&	0.00	&	0.00	&	0.00	&	0.00	&	0.00	&	0.50	&	0.00	&	0.00	&	0.00	&	0.21	&	0.00	&	0.00	\\
6	&	0.01	&	0.00	&	0.00	&	0.00	&	0.00	&	0.00	&	0.00	&	0.00	&	0.00	&	0.00	&	0.00	&	0.03	&	0.85	\\
7	&	0.02	&	0.00	&	0.61	&	0.00	&	0.00	&	0.00	&	0.00	&	0.46	&	0.03	&	0.16	&	0.00	&	0.00	&	0.00	\\
8	&	0.02	&	0.00	&	0.23	&	0.00	&	0.00	&	0.00	&	0.00	&	0.27	&	0.54	&	0.15	&	0.00	&	0.00	&	0.00	\\
9	&	0.00	&	0.00	&	0.00	&	0.00	&	0.00	&	0.00	&	0.42	&	0.02	&	0.00	&	0.00	&	0.00	&	0.00	&	0.00	\\
10	&	0.00	&	0.00	&	0.00	&	0.46	&	0.71	&	0.01	&	0.00	&	0.00	&	0.00	&	0.02	&	0.00	&	0.00	&	0.00	\\
11	&	0.00	&	0.00	&	0.00	&	0.00	&	0.00	&	0.00	&	0.00	&	0.13	&	0.09	&	0.03	&	0.79	&	0.00	&	0.00	\\
12	&	0.03	&	0.56	&	0.01	&	0.00	&	0.02	&	0.00	&	0.00	&	0.01	&	0.01	&	0.04	&	0.00	&	0.00	&	0.00	\\
13	&	0.00	&	0.00	&	0.00	&	0.00	&	0.00	&	0.00	&	0.00	&	0.00	&	0.00	&	0.00	&	0.00	&	0.62	&	0.01	\\
14	&	0.00	&	0.00	&	0.00	&	0.54	&	0.14	&	0.00	&	0.00	&	0.00	&	0.00	&	0.01	&	0.00	&	0.00	&	0.00	\\
15	&	0.12	&	0.37	&	0.09	&	0.00	&	0.00	&	0.00	&	0.00	&	0.01	&	0.00	&	0.15	&	0.00	&	0.02	&	0.00	\\

\label{SC_15_clusters}
\end{longtabu}
\end{spacing}
\end{landscape}

\newpage
\begin{landscape}

\begin{spacing}{1.0}
\small

\begin{longtabu}{p{0.2cm}p{0.8cm}p{0.9cm}p{1.4cm}p{1.3cm}p{1.65cm}p{1.6cm}p{1.4cm}p{2cm}p{2cm}p{1.6cm}p{1.4cm}p{1.4cm}p{1.5cm}}
\caption{Clustering abstracts in the \emph{conflated\_SC\_sampled} dataset with 20 clusters.}\\

\hline\noalign{\smallskip}
\hline\noalign{\smallskip}
&Astro&Astro-M	&Astro \& GeoPhys&BioChem \& MBio&	BioChem \& MBio-M&BioTech \& AMBio-M&EnvSc \& Eco-M&	GeoChem \& GeoPhys&GeoChem \& GeoPhys-M&	Multidisc-\newline iplinary&Oceano-\newline graphy&Physics&Physics-M\\
\hline\noalign{\smallskip}
\endfirsthead

\hline\noalign{\smallskip}
\hline\noalign{\smallskip}
&Astro&Astro-M	&Astro \& GeoPhys&BioChem \& MBio&	BioChem \& MBio-M&BioTech \& AMBio-M&EnvSc \& Eco-M&	GeoChem \& GeoPhys&GeoChem \& GeoPhys-M&	Multidisc-\newline iplinary&Oceano-\newline graphy&Physics&Physics-M\\ 
\hline\noalign{\smallskip}
\endhead
\hline\noalign{\smallskip}
\endfoot

\multicolumn{2}{l}{Trial~1}\\
\hline
1	&	0.01	&	0.00	&	0.23	&	0.00	&	0.00	&	0.00	&	0.00	&	0.25	&	0.08	&	0.07	&	0.00	&	0.00	&	0.00	\\
2	&	0.10	&	0.05	&	0.05	&	0.00	&	0.00	&	0.00	&	0.00	&	0.07	&	0.08	&	0.18	&	0.00	&	0.00	&	0.08	\\
3	&	0.00	&	0.00	&	0.00	&	0.00	&	0.00	&	0.00	&	0.42	&	0.01	&	0.00	&	0.00	&	0.00	&	0.00	&	0.00	\\
4	&	0.02	&	0.00	&	0.10	&	0.00	&	0.00	&	0.00	&	0.00	&	0.05	&	0.29	&	0.06	&	0.00	&	0.00	&	0.00	\\
5	&	0.00	&	0.00	&	0.00	&	0.00	&	0.00	&	0.00	&	0.35	&	0.00	&	0.00	&	0.00	&	0.21	&	0.00	&	0.00	\\
6	&	0.00	&	0.00	&	0.00	&	0.07	&	0.21	&	0.03	&	0.00	&	0.00	&	0.00	&	0.02	&	0.00	&	0.00	&	0.00	\\
7	&	0.53	&	0.00	&	0.00	&	0.00	&	0.00	&	0.00	&	0.00	&	0.00	&	0.00	&	0.06	&	0.00	&	0.00	&	0.01	\\
8	&	0.00	&	0.00	&	0.00	&	0.00	&	0.00	&	0.00	&	0.00	&	0.08	&	0.08	&	0.03	&	0.53	&	0.00	&	0.00	\\
9	&	0.14	&	0.14	&	0.04	&	0.00	&	0.00	&	0.00	&	0.00	&	0.00	&	0.00	&	0.12	&	0.00	&	0.04	&	0.00	\\
10	&	0.00	&	0.00	&	0.00	&	0.34	&	0.25	&	0.01	&	0.00	&	0.00	&	0.00	&	0.00	&	0.00	&	0.00	&	0.00	\\
11	&	0.02	&	0.03	&	0.01	&	0.00	&	0.00	&	0.00	&	0.01	&	0.09	&	0.37	&	0.09	&	0.00	&	0.00	&	0.00	\\
12	&	0.00	&	0.00	&	0.00	&	0.00	&	0.03	&	0.96	&	0.06	&	0.00	&	0.00	&	0.01	&	0.00	&	0.00	&	0.00	\\
13	&	0.00	&	0.00	&	0.00	&	0.59	&	0.51	&	0.00	&	0.00	&	0.00	&	0.00	&	0.02	&	0.00	&	0.00	&	0.00	\\
14	&	0.01	&	0.00	&	0.00	&	0.00	&	0.00	&	0.00	&	0.00	&	0.00	&	0.00	&	0.00	&	0.00	&	0.35	&	0.06	\\
15	&	0.01	&	0.57	&	0.00	&	0.00	&	0.00	&	0.00	&	0.00	&	0.00	&	0.00	&	0.01	&	0.00	&	0.00	&	0.00	\\
16	&	0.02	&	0.00	&	0.01	&	0.00	&	0.00	&	0.00	&	0.00	&	0.00	&	0.00	&	0.01	&	0.00	&	0.59	&	0.01	\\
17	&	0.02	&	0.00	&	0.56	&	0.00	&	0.00	&	0.00	&	0.00	&	0.39	&	0.02	&	0.12	&	0.00	&	0.00	&	0.00	\\
18	&	0.12	&	0.21	&	0.01	&	0.00	&	0.00	&	0.00	&	0.00	&	0.02	&	0.08	&	0.18	&	0.00	&	0.00	&	0.00	\\
19	&	0.01	&	0.00	&	0.00	&	0.00	&	0.00	&	0.00	&	0.00	&	0.00	&	0.00	&	0.00	&	0.00	&	0.01	&	0.84	\\
20	&	0.00	&	0.00	&	0.00	&	0.00	&	0.00	&	0.00	&	0.17	&	0.03	&	0.00	&	0.01	&	0.26	&	0.00	&	0.00	\\
\hline\noalign{\smallskip}
\multicolumn{2}{l}{Trial~2}\\
\hline																											
1	&	0.00	&	0.00	&	0.00	&	0.19	&	0.18	&	0.02	&	0.00	&	0.00	&	0.00	&	0.00	&	0.00	&	0.00	&	0.00	\\
2	&	0.02	&	0.00	&	0.76	&	0.00	&	0.00	&	0.00	&	0.00	&	0.51	&	0.05	&	0.17	&	0.00	&	0.00	&	0.00	\\
3	&	0.00	&	0.00	&	0.00	&	0.00	&	0.00	&	0.00	&	0.40	&	0.02	&	0.00	&	0.00	&	0.00	&	0.00	&	0.00	\\
4	&	0.00	&	0.00	&	0.00	&	0.16	&	0.21	&	0.01	&	0.00	&	0.00	&	0.00	&	0.00	&	0.00	&	0.00	&	0.00	\\
5	&	0.01	&	0.00	&	0.00	&	0.00	&	0.00	&	0.00	&	0.00	&	0.00	&	0.00	&	0.00	&	0.00	&	0.00	&	0.61	\\
6	&	0.10	&	0.01	&	0.00	&	0.00	&	0.00	&	0.00	&	0.00	&	0.01	&	0.02	&	0.05	&	0.00	&	0.15	&	0.16	\\
7	&	0.00	&	0.00	&	0.00	&	0.00	&	0.00	&	0.00	&	0.00	&	0.00	&	0.00	&	0.00	&	0.00	&	0.14	&	0.23	\\
8	&	0.01	&	0.62	&	0.00	&	0.00	&	0.00	&	0.00	&	0.00	&	0.00	&	0.00	&	0.02	&	0.00	&	0.00	&	0.00	\\
9	&	0.00	&	0.00	&	0.00	&	0.09	&	0.36	&	0.00	&	0.00	&	0.00	&	0.00	&	0.02	&	0.00	&	0.00	&	0.00	\\
10	&	0.00	&	0.00	&	0.00	&	0.35	&	0.06	&	0.00	&	0.00	&	0.00	&	0.00	&	0.00	&	0.00	&	0.00	&	0.00	\\
11	&	0.00	&	0.00	&	0.00	&	0.00	&	0.00	&	0.00	&	0.01	&	0.03	&	0.00	&	0.01	&	0.32	&	0.00	&	0.00	\\
12	&	0.00	&	0.00	&	0.00	&	0.00	&	0.00	&	0.00	&	0.50	&	0.00	&	0.00	&	0.00	&	0.17	&	0.00	&	0.00	\\
13	&	0.18	&	0.13	&	0.02	&	0.00	&	0.00	&	0.00	&	0.00	&	0.06	&	0.21	&	0.30	&	0.00	&	0.00	&	0.00	\\
14	&	0.00	&	0.00	&	0.00	&	0.21	&	0.10	&	0.02	&	0.00	&	0.00	&	0.00	&	0.01	&	0.00	&	0.00	&	0.00	\\
15	&	0.02	&	0.00	&	0.01	&	0.00	&	0.00	&	0.00	&	0.00	&	0.00	&	0.00	&	0.01	&	0.00	&	0.68	&	0.00	\\
16	&	0.63	&	0.04	&	0.04	&	0.00	&	0.00	&	0.00	&	0.00	&	0.00	&	0.00	&	0.16	&	0.00	&	0.02	&	0.01	\\
17	&	0.00	&	0.00	&	0.00	&	0.00	&	0.08	&	0.96	&	0.08	&	0.00	&	0.00	&	0.01	&	0.00	&	0.00	&	0.00	\\
18	&	0.01	&	0.00	&	0.03	&	0.00	&	0.00	&	0.00	&	0.01	&	0.27	&	0.53	&	0.16	&	0.00	&	0.00	&	0.00	\\
19	&	0.00	&	0.00	&	0.00	&	0.00	&	0.00	&	0.00	&	0.00	&	0.03	&	0.03	&	0.01	&	0.50	&	0.00	&	0.00	\\
20	&	0.03	&	0.20	&	0.14	&	0.00	&	0.00	&	0.00	&	0.00	&	0.05	&	0.17	&	0.07	&	0.00	&	0.00	&	0.00	\\
\hline\noalign{\smallskip}
\multicolumn{2}{l}{Trial~3}\\
\hline																						
1	&	0.00	&	0.00	&	0.00	&	0.42	&	0.30	&	0.00	&	0.00	&	0.00	&	0.00	&	0.01	&	0.00	&	0.00	&	0.00	\\
2	&	0.00	&	0.00	&	0.00	&	0.00	&	0.00	&	0.00	&	0.00	&	0.00	&	0.00	&	0.00	&	0.00	&	0.01	&	0.70	\\
3	&	0.00	&	0.00	&	0.00	&	0.00	&	0.00	&	0.00	&	0.41	&	0.02	&	0.00	&	0.00	&	0.00	&	0.00	&	0.00	\\
4	&	0.00	&	0.00	&	0.00	&	0.35	&	0.26	&	0.02	&	0.00	&	0.00	&	0.00	&	0.00	&	0.00	&	0.00	&	0.00	\\
5	&	0.04	&	0.19	&	0.01	&	0.00	&	0.05	&	0.00	&	0.01	&	0.04	&	0.01	&	0.12	&	0.00	&	0.00	&	0.00	\\
6	&	0.01	&	0.00	&	0.00	&	0.00	&	0.00	&	0.00	&	0.00	&	0.00	&	0.00	&	0.00	&	0.00	&	0.43	&	0.01	\\
7	&	0.14	&	0.07	&	0.11	&	0.00	&	0.00	&	0.00	&	0.00	&	0.01	&	0.00	&	0.16	&	0.00	&	0.04	&	0.00	\\
8	&	0.01	&	0.00	&	0.25	&	0.00	&	0.00	&	0.00	&	0.00	&	0.16	&	0.34	&	0.08	&	0.00	&	0.00	&	0.00	\\
9	&	0.00	&	0.00	&	0.00	&	0.24	&	0.32	&	0.00	&	0.00	&	0.00	&	0.00	&	0.02	&	0.00	&	0.00	&	0.00	\\
10	&	0.02	&	0.00	&	0.00	&	0.00	&	0.00	&	0.00	&	0.00	&	0.01	&	0.01	&	0.01	&	0.00	&	0.06	&	0.28	\\
11	&	0.00	&	0.00	&	0.00	&	0.00	&	0.00	&	0.00	&	0.50	&	0.00	&	0.00	&	0.00	&	0.20	&	0.00	&	0.00	\\
12	&	0.00	&	0.00	&	0.00	&	0.00	&	0.00	&	0.00	&	0.00	&	0.02	&	0.02	&	0.00	&	0.80	&	0.00	&	0.00	\\
13	&	0.17	&	0.01	&	0.02	&	0.00	&	0.00	&	0.00	&	0.00	&	0.07	&	0.29	&	0.29	&	0.00	&	0.00	&	0.00	\\
14	&	0.00	&	0.32	&	0.02	&	0.00	&	0.00	&	0.00	&	0.00	&	0.00	&	0.00	&	0.01	&	0.00	&	0.00	&	0.00	\\
15	&	0.02	&	0.00	&	0.59	&	0.00	&	0.00	&	0.00	&	0.00	&	0.44	&	0.03	&	0.14	&	0.00	&	0.01	&	0.00	\\
16	&	0.55	&	0.00	&	0.00	&	0.00	&	0.00	&	0.00	&	0.00	&	0.00	&	0.00	&	0.07	&	0.00	&	0.02	&	0.01	\\
17	&	0.00	&	0.00	&	0.00	&	0.00	&	0.08	&	0.98	&	0.06	&	0.00	&	0.00	&	0.01	&	0.00	&	0.00	&	0.00	\\
18	&	0.01	&	0.00	&	0.00	&	0.00	&	0.00	&	0.00	&	0.02	&	0.22	&	0.30	&	0.07	&	0.00	&	0.00	&	0.00	\\
19	&	0.00	&	0.00	&	0.00	&	0.00	&	0.00	&	0.00	&	0.00	&	0.00	&	0.00	&	0.00	&	0.00	&	0.43	&	0.00	\\
20	&	0.01	&	0.42	&	0.00	&	0.00	&	0.00	&	0.00	&	0.00	&	0.00	&	0.00	&	0.01	&	0.00	&	0.00	&	0.00	\\

\label{SC_20_clusters}
\end{longtabu}
\end{spacing}
\end{landscape}

\end{document}